# A Review on Energy Consumption Optimization Techniques in IoT Based Smart Building Environment

**Abdul Salam Shah [1], Haidawati Nasir [1,*], Muhammad Fayaz [2], Adidah Lajis [1], Asadullah Shah [3]**

[1] Department of Computer Engineering, University of Kuala Lumpur (UniKl-MIIT), 1016 Jalan Sultan Ismail 50250 Kuala Lumpur, Malaysia; shahsalamss@gmail.com (A.S.S.); adidahl@unikl.edu.my (A.L.)
[2] Department of Computer Engineering, Jeju National University, Jejusi 63243, Jeju Special Self-Governing Province, Korea; hamaz_khan@yahoo.com (M.F.);
[3] Department of Information Systems, Kulliyyah of ICT, International Islamic University Malaysia (IIUM), Gombak Campus, Kuala Lumpur, Malaysia; asadullah@iium.edu.my (A.S.);
* Correspondence: haidawati@unikl.edu.my (H.N.); Tel.: +60-18-982-6987

**Abstract:** In recent years, due to the unnecessary wastage of electrical energy in residential buildings, the requirement of energy optimization and user comfort has gained vital importance. In the literature, various techniques have been proposed addressing the energy optimization problem. The goal of each technique was to maintain a balance between user comfort and energy requirements such that the user can achieve the desired comfort level with the minimum amount of energy consumption. Researchers have addressed the issue with the help of different optimization algorithms and variations in the parameters to reduce energy consumption. To the best of our knowledge, this problem is not solved yet due to its challenging nature. The gap in the literature is due to the advancements in the technology and drawbacks of the optimization algorithms and the introduction of different new optimization algorithms. Further, many newly proposed optimization algorithms which have produced better accuracy on the benchmark instances but have not been applied yet for the optimization of energy consumption in smart homes. In this paper, we have carried out a detailed literature review of the techniques used for the optimization of energy consumption and scheduling in smart homes. The detailed discussion has been carried out on different factors contributing towards thermal comfort, visual comfort, and air quality comfort. We have also reviewed the fog and edge computing techniques used in smart homes.

## 1. Introduction

The world of information and communication technology is advancing with the addition of new sensing and communication technologies, to connect from anyplace, anytime and anything. This type of connectivity is known as the Internet of Things (IoT) [1,2]. The security, connectivity, privacy and standard procedures for communication in the IoT based network is the biggest concern nowadays [3]. Researchers are trying to resolve the concern so that the IoT based networks can be successfully utilized in the real environment [4]. The electricity consumption prediction and optimization in residential building is also the concern of researchers and scientists to make IoT based smart home systems successful. The traditional methods used for energy management are forecasting based on statistical analysis and machine learning approaches applied to data of energy consumption gathered from electricity meters [5]. However, with traditional methods, the hourly energy consumption prediction is not possible. If we use data collected from digital meters and apply machine learning techniques on them it will be possible to predict hourly energy consumption and based on predicted energy consumption; we can optimize energy consumption, to avoid the wastage of precious energy resources. Regarding the energy consumption prediction, much work has already been done based on yearly, monthly, weekly, daily, and hourly consumption using Mean Absolute Error (MAE), Mean Squared Error (MSE) and Root Mean Squared Error (RMSE) as evaluation parameters [6-8].

The power consumption prediction remained a concern for the power generation companies due to the growing demand for energy caused by the rapid increase in the world population. The scientists believe that if the energy consumption has not controlled, it may lead to the shortfall of the energy after a few years. There are two options to deal with the shortfall of energy; 1) producing more energy, 2) minimizing the consumption of already available energy resources and reducing the wastage. The production of energy is a very costly solution to the problem which requires a lot of time and resources, but on the other hand, by taking some pre-emptive measures, the energy consumption minimization can be achieved [9]. Since the last few decades, much research has been done by the researchers on energy consumption prediction and optimization. The energy prediction is first stepping to take for the optimization of energy consumption. Based on previously consumed energy, we must predict the energy consumption of the next hour, month or year. The concept of energy optimization is helpful in the smart homes where the devices continuously consume the same amount of energy, but if we implement the concept of optimization in the smart homes, it will supply the exact amount of power to the appliances [10]. The optimization technique performs based on external conditions of the room like temperature, illumination, humidity, air flow, air quality and so forth. The satisfaction of the residents of a building is an important factor due to which the smart homes and their energy consumption optimization are becoming an exciting topic for scientists and researchers. The researchers have proposed different techniques based on some optimization algorithms to tackle the challenges of energy management and improve the comfort index of residents of smart buildings. The ant colony optimization and fuzzy logic were used for the optimization of energy consumption, but still, there is a gap (regarding the cooling scenario), and resident comfort index needs to improve to some extent with less power consumption [11]. The major issue with the already available techniques is the dependency on the geographic area of the system and differentiation in environmental conditions. The maintenance of the balance between the comfort index and power consumption is also a significant issue [12,13]. The rule design according to the environmental conditions in fuzzy logic is another issue [14]. The researchers aim to increase the resident's comfort index and reduce the power consumption through optimization techniques and rule-based optimization. By predicting indoor environment parameters, the reduction in power consumption is possible. For the justification of carrying out this research, it is essential to understand the terms provided hereafter.

To the best of knowledge, no review paper has covered the RQs. To answer the RQs, we have reviewed the literature on optimization algorithms used for the optimization of energy consumption and scheduling. The optimization algorithms are based on the approach to find minimum values of mathematical functions. They are used to evaluate design trade-offs, to assess control systems, and to find patterns in data. The solution of optimization problem consists of a reduction of a complex problem into the more straightforward problem and then solving each straightforward problem and using its solution to solve the next problem [15]. The mathematical optimization is the selection of best elements from some set of available alternatives. The optimization algorithms maximize or minimize the real function by selecting an input value from a given set of values and based on that value estimate the new value of the given function. Based on continuous and discrete values the optimization problems are divided into two problems [16]. It is required to find optimal parameters using optimization techniques in energy management systems. The parameters are divided into two categories the environmental conditions in the smart home area as current indoor parameters and the demanded parameters by a resident of a smart home as user set parameters [17]. These parameters are temperature, illumination, humidity airflow, air quality and so forth. The error difference used for the optimization is the difference between current parameters and user-set parameters. The minimum error difference means minimum power consumption, so the target of techniques is to reduce the error difference to the minimum level that will lead to minimizing power consumption. Researchers have already used optimization techniques to achieve minimum error difference, but still, further reduction in error difference is possible. It has been observed from the literature review that researchers have tried to solve the optimization problem with two different ways, i.e., 1) scheduling, 2) optimization.

The review of the literature revealed that many researchers have considered the optimization of the overall building or house, which results in less energy saving [8]. There is a gap in the literature to optimize the individual areas of the house as it will save more energy. Further, in the upper portion of the house, the temperature remains higher as compared to the lower portion so the division of overall optimization into different subparts will save more energy. In literature, a lot of research regarding energy consumption optimization has been done, but still, the work is in progress to make the systems smarter regarding energy optimization [11].

To the best of knowledge, no review paper has covered the literature on energy consumption optimization and energy consumption scheduling in smart homes. Therefore, this study aims to carry out a literature review on energy consumption optimization and scheduling in order to:

(a) Identify algorithms/techniques used for energy consumption optimization and energy consumption scheduling in smart homes.

(b) Identify edge and fog computing techniques used in smart homes.

(c) Identify comfort index parameters in smart homes.

(d) Identify the technologies used in smart homes.

(e) Present a synthesis of empirical evidence found in (a), (b), (c) and (d).

Although both techniques (optimization and scheduling) are reviewed, the focus of this paper is on optimization. Therefore, the more detailed critical analysis focuses on optimization as compared to scheduling techniques.

The organization of the rest of the paper is as; section 2 presents a review methodology, results are provided in section 3. Section 4 contains detailed discussion, section 5 contains a conclusion, and finally, section 6 presents the future work directions of the study.

## 2. Review Methodology

### *2.1. Research Questions*

The definition of research questions (RQs) is necessary for any mapping and systematic studies. The research questions of this study are defined as:

RQ1: Which algorithms/techniques and parameters have been used for energy consumption optimization in smart homes?

RQ2: Which algorithms/techniques, parameters, and pricing schemes have been used for energy consumption scheduling in smart homes?

RQ3: How the edge and fog computing techniques are used in smart homes?

RQ4: What are the technologies used in smart homes for the connectivity of devices?

RQ5: What are the different comfort index parameters in smart homes?

### *2.2. Searching for Literature*

The literature has been searched from the Web of Science, IEEE Xplore, Google Scholar, and Scopus. These databases cover topics related to energy optimization and scheduling and have enough literature. The searching criteria base on different factors as mentioned in the QRs. The search strings defined in Table 1, have been identified and used to search each database using Mendeley, Google Scholar, and Endnote.

**Table 1**. Search strings for searching for the literature

| Sr. No. | Search Strings | Starting Year | End Year |
|---|---|---|---|
| 1. | Genetic algorithms for energy optimization | 1996 | 2018 |
| 2. | Energy optimization in smart building | 1996 | 2018 |
| 3. | Edge computing in smart building | 2009 | 2018 |
| 4. | Fog computing in smart building | 2009 | 2018 |
| 5. | Energy scheduling in smart building | 2009 | 2018 |
| 6. | Internet of Things (IoT) in smart building | 2009 | 2018 |

### *2.3. Inclusion/Exclusion of Literature*

The selection of the articles has been carried out based on the consideration of comfort index and energy optimization. The focus is on genetic algorithm-based techniques using a proportional–integral–derivative (PID) and fuzzy logic controllers. Further, we have considered the articles which have focused on reducing cost and saving energy. The fog and edge computing are relatively new domains regarding energy optimization; therefore, we have selected the maximum possible articles for study based on its application in smart homes. The detailed inclusion and exclusion criteria are defined in Table 2.

**Table 2.** Inclusion-exclusion criteria of literature

| Optimization Techniques | | Scheduling, Fog, Edge, Techniques | |
|---|---|---|---|
| **Inclusion Criteria** | **Exclusion Criteria** | **Inclusion Criteria** | **Exclusion Criteria** |
| Publication date 1996 - 2018. | Published Pre-1996. | Publication date 2009 - 2018. | Published Pre-2009. |
| Any geographical location. | Patent. | Any geographical location. | Patent. |
| English language. | Non-English. | English language. | Non-English. |
| Grey literature Reports, standards. | Dissertation/Thesis. | Grey literature Reports, standards. | Dissertation/Thesis. |
| The articles published in peer-reviewed journals of Web of Science, Scopus, IEEE Xplore and conference articles/proceedings answering defined research RQs. | The articles regarding the energy-efficient design of buildings have been excluded. | The literature has been included based on the scheduling techniques and different optimization algorithms as defined in RQs. | The articles outside of the scope of scheduling have been excluded. |

## 3. Results

The flow diagram of the inclusion/ exclusion of the articles for literature review can be seen in Figure 1. [18]. The following number of articles have been retrieved: 'edge computing in smart building' (131), 'fog computing in smart building' (122), 'genetic algorithms for energy optimization' (1940), 'energy optimization in smart building' (585), 'energy scheduling in smart building' (279), and 'Internet of Things (IoT) in smart building' (708). Total of 3765 articles has been retrieved from database searches (Web of Science, IEEE Xplore, Scopus), further 35 articles have been identified using Google Scholar. Mendeley and Endnote have been used to identify and delete duplicates because it can identify and separate the duplicates automatically. After duplicate removal of 100 articles total 2700 left. After title and abstract review, 3400 articles have been excluded because they were not relevant to the topic, editorials, reports, irrelevant outcome measure. We have conducted the full article review on the remaining 300 articles, and 201 were excluded due to non-consideration of comfort index, scheduling in other domains, repeated studies and focus on building structure rather than optimization techniques.

The other main reason of exclusion of the articles regarding fog and edge computing was the general implementation of techniques in IoT rather than smart buildings. We have applied forward and backward snowballing technique on the 99 articles for the further literature search on the reference list of the articles up to 30 iterations in both directions or to the year of study under consideration, i.e., 1996 for the energy optimization techniques and 2009 for scheduling and fog, edge techniques. For the most recently published articles, the forward snowballing has been carried out until the last citation for the identification of further articles. The initial seed for the forward snowballing was the oldest paper under study limit. Similarly for backward snowballing the latest articles of the year 2018 has been selected. The priority has given to the articles published in recent years. We have identified 16 articles using snowballing, and finally, 115 articles have qualified for the review. The detail of the included articles is provided in Appendix A Table A1. The remaining articles, reports, standards in the study are the supporting material. The articles have been shortlisted based on the title, abstract, and keywords in comparison with the selection criteria. The articles which have not fulfilled the criteria and having the same nature of study have been excluded with noting their reasons.

The relevant information from the articles related to energy optimization has been extracted and reported. The criteria of the relevant information for each section is different based on the nature of the studies. Further, we have selected 36 articles for the complete analysis for the optimization techniques and 20 for scheduling techniques. From the total of 115 articles selected for the review detailed review of 56 articles has been carried out, the remaining articles have been formally reviewed, and relevant information has been extracted to answer the RQs. The complete detail of the articles included in the review for the optimization techniques can be seen in Appendix B Table A2, and for scheduling in Appendix C Table A3.

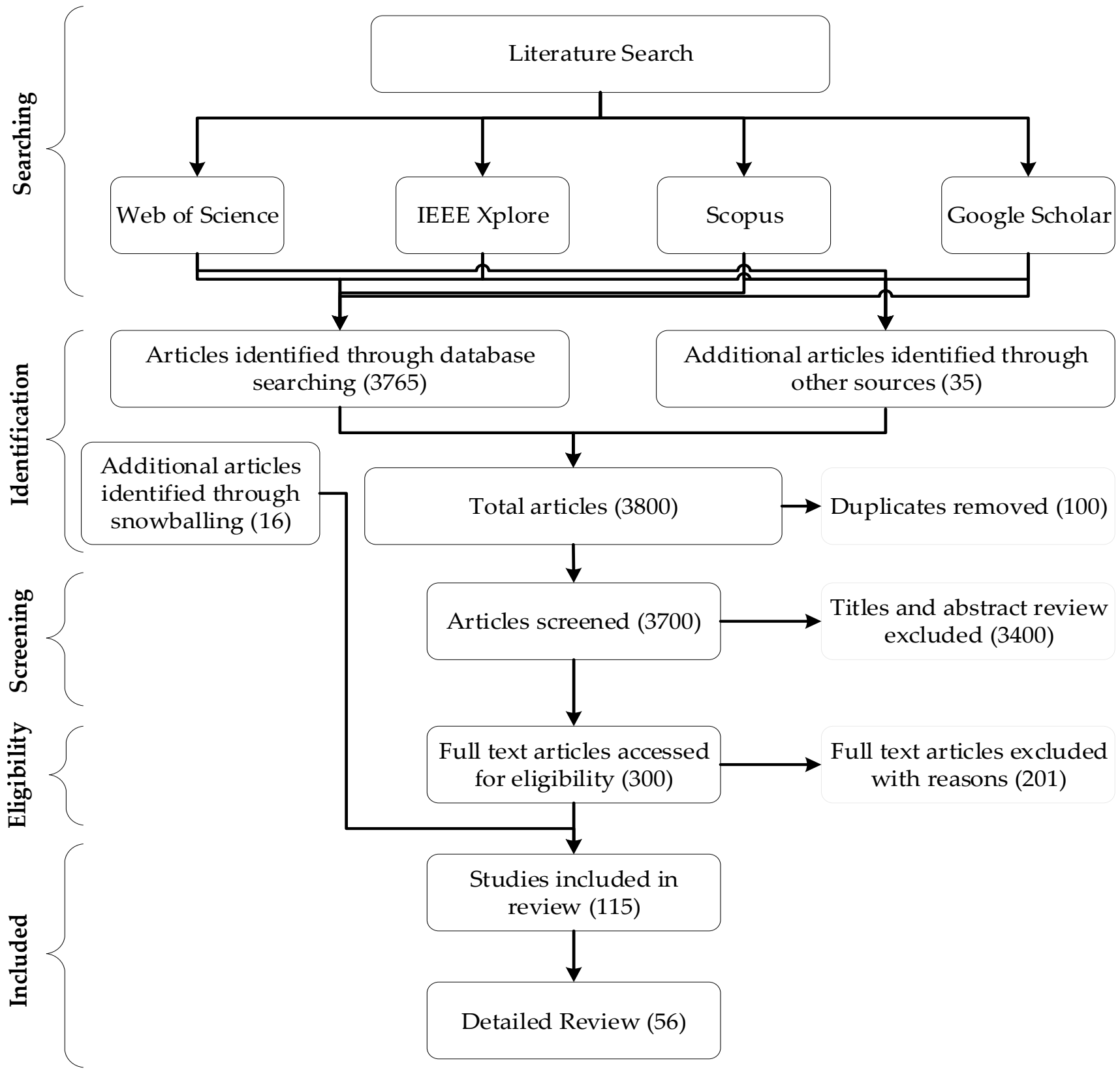

**Figure 1.** The flow diagram of the articles for the literature review

## 4. Discussion

### *4.1. Research Question 1: RQ1: Which algorithms/techniques and parameters have been used for energy consumption optimization in smart homes?*

To answer the RQ1, we have extracted following relevant information from each article: 1) algorithm/technique, 2) focus area, 3) energy prediction, 4) energy optimization, 5) thermal comfort, 6) visual comfort, 7) air quality comfort, 8) sound comfort, 9) air quality, 10) temperature, 11) humidity, 12) illumination, 13) airflow, 14) heat radiation/flow, 15) cloth thermal insulation, 16) metabolic rate, 17) water vapor pressure, 18) single user, 19) multi-user, 20) individual room, 21) external environment. Appendix B, Table A2 discusses detailed information about these parameters. For the calculation of frequencies of each parameter and algorithms/techniques, comfort index, and frequency of publications from each year we have carried out systematic mapping using bubble technique reported in [19].

4.1.1. Systematic Mapping of Optimization Techniques

The data of Appendix B, Table A2, has been used for the systematic mapping same can be seen in Figure 2 [19]. The other relevant information regarding the percentage of each technique, parameters, and so forth has been saved in Excel sheet and elaborated in Figure 2 for the optimization techniques. The frequency of the articles included from each year is also provided with their percentage for better understanding and systematic extension. The definition of abbreviations used in the parameter portion of the map is provided in the note portion of Appendix B, Table A2 and the remaining are provided as: particle swarm optimization (PSO), genetic algorithm (GA), artificial neural network (ANN), decision support model (DSM), linear reinforcement learning controller (LRLC), heuristic system identification (HIS), model-based predictive control strategy (MBP), Markov decision problems (MDP), ant colony optimization algorithm (ACO), multinomial logistic regression (MLR), bat algorithm (BAT), and artificial bee colony (ABC).

The map in Figure 2 provides complete detail of the 36 articles included in the study for the optimization techniques. We have observed that most of the authors have given priority to the thermal comfort hence 33 (91.66%) out of 36 articles have considered the thermal comfort. Moreover, for the thermal comfort 32 (88.88%) articles have considered temperature, 11 (30.55%) have considered humidity, 7 (19.44%) airflow and minor consideration of other parameters. Total 13 (36.11%) articles have considered the external environmental parameters. If we consider the frequency of prediction optimization techniques used by the authors then, 32 (88.88%) articles have focused on energy optimization, 8 (22.22%) on energy prediction. The other most important factor is the number of users focused by studies for satisfaction and comfort index consideration. Mostly the authors have followed standard values of comfort index and focused on the standards rather than the preferences of everyone. Regarding the user preference 28 (77.77%) articles have considered the preference of a single user according to the standard set of different comforts. Only 2 (5.55%) have considered the multiple user comfort index, and 8 (22.22%) articles have considered the individual room for the energy optimization rather than the complete building.

The other most important comfort is the visual comfort which has been considered by 19 (52.77%) articles, considering the illumination by 19 (52.77%) as the visual parameter. The air quality comfort has been considered by 16 (44.44%) articles using $CO_2$ 16 (44.44%) as air quality parameter.

Overall 34 (94.44%) articles have considered the comfort index, 27 (75.00%) energy saving, 1 (2.77%) occupant behavior, 1 (2.77%) learning to control, and 1 (2.77%) multiple user comfort index. Regarding the optimization algorithms, the genetic algorithm has been used by 10 (27.77%) articles for comfort index, 7(19.44%) for energy saving, and 1(2.77%) for learning to control. The other most prominent algorithm was the artificial neural network and its variations, which has been used by 6 (16.66%) articles for comfort index, and 4 (11.11%) for energy saving. Furthermore, particles swarm optimization algorithm has been used by 2 (5.55%) articles for the comfort index and 3 (8.33%) for the energy saving. Very few authors have used the remaining algorithms for the energy saving and comfort index management.

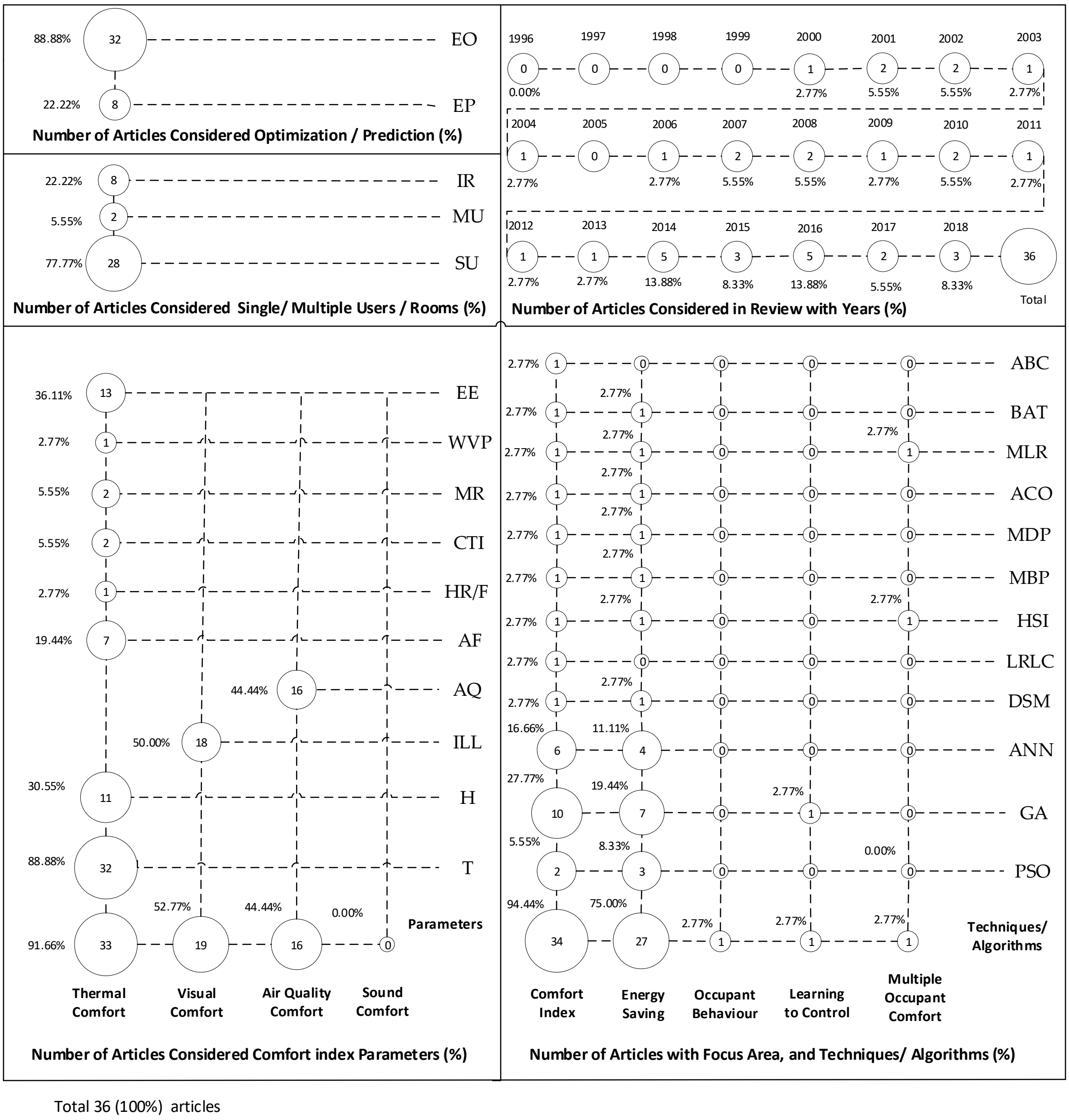

**Figure 2.** Visualization of a systematic map of optimization techniques

#### 4.1.2. Algorithms / Techniques used for the Energy Optimization

The research in the area of energy management systems has started decades back. For the energy saving and optimization, different researchers have used various methodologies to achieve the target. For the detailed comparative analysis kindly refer Appendix B Table A2. For the tuning of proportional integral and derivative (PID), in the heating ventilation and air conditioning system genetic algorithm was used by [17]. The performance of the system was measured using overshoot, setting time and mean squared error [17].

The trend has started using conventional control systems based on classical controllers since last few decades [20]. With the advancement in technology, the demand for the change has grown due to the overshooting temperature problem with these controllers. The other issues include energy efficiency and user preference due to which the conventional models were not suitable. For the solution of temperature overshooting problem proportional–integral–derivative (PID) controllers were introduced [20]. The PID controllers also have some disadvantages like the selection of the wrong algorithm, improper tuning, over or under-filtering, incorrect configuration of the control strategy, spanning and scaling, and scan related issues.

To overcome the problems of PID controllers, fuzzy controllers have achieved vital importance and using a genetic algorithm for the optimization, and fuzzy logic for controlling a method was introduced by [21]. They have focused on increasing the comfort index and minimizing energy consumption. The technique has considered the user preference to improve the comfort index. The performance of the controllers was measured using overshoot, setting time, and mean squared error [21]. A predictive control method using a genetic algorithm for the VAV air conditioning system was introduced by [22]. The method based on the PID controller for the control of temperature based on self-tuning. The process does not consider the single or multi-user comfort index the focus is just the energy efficiency.

A method for the comfort index and energy consumption optimization using fuzzy PID, fuzzy PD, and adaptive fuzzy PD was introduced by [23]. They have considered the thermal, visual and air quality comfort. The primary objective of the control system was the satisfaction of user preferences, avoid the overshooting and energy waste, and monitoring of energy consumption. They have given preference to the residents of the building. The rule base was accordingly designed to overcome the issue of overshooting. The other method for the adaptive controllers has been proposed in [24] which based on the neural network. The model predicts outdoor parameters using ANN, but it does not consider the user preferences and comfort index of the residents. The authors have ignored the factors like illumination and air quality and considered the temperature only.

For the optimum payoff characteristics between energy cost and thermal comfort, multi-objective genetic algorithm (MOGA) was used by [25]. They have considered the comfort index of the resident of the building and found the pricing of the energy after some time intervals. They have not discussed the visual and air quality comfort. The optimized fuzzy controller has been proposed to improve comfort index and reduce energy consumption by [26]. They had used ON/OFF method when the indoor illumination reached the desired level the lighting turns OFF automatically. They have introduced the user parameters to the system through a smart card [26]. The system was not compatible with the particular changes in the environmental conditions, like the change in temperature due to the specific opening of windows and so forth [26].

A comparative study of the different fuzzy controllers like fuzzy P, fuzzy PID, fuzzy PI, fuzzy PD, and adaptive fuzzy PD has been carried out in [27]. The performance criteria of the model were energy consumption and response performance; they have not considered the user preferences for the performance of the system [27]. They have addressed the overshooting problem in the methodology which is the main reason for the wastage of energy in the residential buildings [27]. The other comprehensive study of the artificial neural network based energy management techniques, fuzzy and adaptive neuro-fuzzy inference systems have been carried out in [28]. The non-linear characteristics are difficult to be monitored and controlled in HVAC systems. For the solution of the issue, a fuzzy control system for HVAC was proposed by [29]. The modeling of the indoor and outdoor environment has been avoided using an adaptive fuzzy controller. They have not considered the user involvement for the comfort index [29].

Fong et al., in [30] proposed an energy efficient evolutionary programming-based metaheuristic approach according to the HVAC system needs for the optimization of energy consumption. Similarly, Lah et al., in [31] used a fuzzy logic controller for the automatic roller blind for illumination control inside the building. The technique focusses to maximize the use of daylight for internal illumination and reduce the energy consumption of artificial lighting [31]. An intelligent decision support model for the building energy management system has been proposed by [32]. It can deal with the diagnosis of the internal environment of the building and reduce energy consumption. The system can be centrally operated for the monitoring of energy consumption in buildings and give commands to actuators based on the rules in the knowledge base [32].

For the comfort index management, a reinforcement learning based method was described by [33], using PPD index to measure the comfort index. They have used linear reinforcement learning controller (LRLC) and fuzzy PD controller. The LRLC faces some difficulties like turn on the wrong system like cooling instead of heating [33]. There is a need for improvement in these methods so that they can adopt the changes of building accordingly and react as per changes [33].

Liang and Du, in [34] have designed a comfort control system using human learning combined with power minimization strategies. The user comfort zones have been tuned using human learning approach which was used in hotels for the short term occupancy by [35]. The studies on thermal comfort in HVAC system using predicted mean value (PMV) method has been carried out by [36]. The proposed strategies focus on improving thermal comfort and minimizing energy consumption. The other major thermal comfort index is least enthalpy estimator (LEE) proposed by [37]. A control algorithm for the indoor environment including thermal, air quality, and visual comfort has been presented by [38].

Moon and Kim, in [39] proposed artificial neural network based model for the thermal comfort management in the buildings. They have considered the temperature, humidity and predicted mean vote (PMV) for the thermal comfort due to the importance of the humidity. Two separate predictive control strategies have been used to deal with the overshoot and undershoot of thermal conditions. The energy saving with the proposed method was not satisfactory due to certain constraints, but the technique has maintained better comfort index. A genetic algorithm has been used for the improvement of the fuzzy rules matrix and membership function for the design of the adaptive fuzzy logic controller by [40]. The fuzzy logic controller has been designed to control the air handling unit of the HVAC system [40]. An intelligent coordinator control system based on master and slave concept was proposed by [41]. The tuning of the parameters of the fuzzy logic controller was carried out using a genetic algorithm.

A multi-agent system for the comfort and building management systems have been proposed by [42]. The model considers the occupancy as well as occupant preferences at the same time and reduces the energy consumption [42]. A fuzzy controller for the HVAC systems based on genetic algorithm has been designed by [43]. The method has focused on the rate of the water and steam in the air handling unit of building and improved the steady-state error, rise time and setting time as compared to traditional fuzzy controllers. Authors in [44] have used the knowledge-based approach for the selection of user indoor environmental preference in an HVAC system. The transformation of multi-objective optimization problem of comfort index and energy optimization into a scalar optimization problem has been carried out using a genetic algorithm and the neural network for the optimization and modeling of the HVAC system in buildings by [45].

Scherer et al., in [46] proposed distributed model predictive (DMPC) for energy management. The model-based predictive controllers have the capability of predicting the room weather forecasting, occupancy, and other related information. The second part of the technique focuses on the implementation of the system in the building's HVAC system for thermal comfort. An intelligent energy efficient agent block function of automation to deal with the heterogeneous energy management systems has been proposed by [47]. The software entity 2eA-FB has the capability of surfing, and reasoning to communicate with the other devices. A zero energy house has been designed to accommodate the thermal, and visual comfort and optimization of energy consumption in the building by [48].

Nagy et al., in [49] have proposed an adaptive lighting control strategy in buildings to improve comfort index and reduce energy consumption. The control scheme uses the data of motion sensors to control the lighting inside the building and turn on off the lights. A framework for the big data analytics and energy saving support system to predict the future energy consumption using previous consumption data and efficiently utilize the energy resources has been proposed by [50]. The system can handle the scheduling of appliances for the energy saving in buildings. Another energy saving controller to adjust the intensity of LED as per input data automatically has been introduced by [51].

Ant colony optimization algorithm has been implemented to achieve the energy saving and improvement in comfort index [52]. Delgarm et al., in [53] used mono and multi-objective particle swarm optimization (MOPSO), for the energy efficiency of the individual room of building regarding heating, cooling and lighting energy consumption. Shaikh et al., in [54] [55] have developed the multiagent control system using a multi-objective genetic algorithm for the energy management inside the building. The approach based on the agents in which master agent is responsible for controlling the parallel agents. The thermal, visual and air quality comfort, has been considered and fuzzy logic used for the required output power. Zheng et al., in [56] summarized the challenges and opportunities regarding zero energy buildings to reduce energy consumption. The influence of the occupant behavior on energy consumption in a building has been discussed by [57]. The study focused on modifying the occupant behavior according to the building environment and helping to design the buildings according to the occupant behavior.

Park et al., in [58] introduced a reinforcement learning based occupant centered controller for the lighting. The LightLearn utilize the occupancy information, switch position, and light information to adjust the lighting according to environmental conditions and user preferences. Multi-occupant comfort methodology for the HVAC system in the buildings using the softmax regression algorithm has been proposed by [59]. The focus was to accommodate multiple occupants and achieve maximum energy saving at the same time. Putra in [60] studied the effect of thermal conditions on the occupant's performance and satisfaction in an office environment. He has used Multiphysics software for the modeling simulation and visualization of the environmental quality. Ain et al., in [61] have used a fuzzy inference system for the comfort index management in smart homes. Mamdani has performed better in hot weather as compared to the Sugeno fuzzy inference system which has proved better results in cold weather.

The trend has moved towards the predictive models, and the neural network-based model was proposed by [62]. Fuzzy logic takes the forecasted environmental parameters from the neural network and gives output accordingly. They have not preferred to consider the user set parameters and focused on the ASHRAE recommended temperature [63] [64]. Authors have focused on the optimal selection of parameters of the model for the accurate prediction. Following the methodology of [62], Collota et al. in [65] used soft computing techniques including a dynamic fuzzy controller and artificial neural network using real variables mapped with a non-linear function for the maximization of comfort index. In the proposed technique, the overfitting problem has been solved using the user's feedback through which the rules have been changed dynamically using Gaussian membership function.

Genetic algorithm and fuzzy logic as a controller of the temperature, illumination and air quality has been used by [66]. A prediction component has also been included in the model using a Kalman filter. The objective of the technique was to improve the comfort index and decrease energy consumption. Wahid et al., [67] carried experimentation for the improvement of comfort index and energy consumption reduction in the residential building using artificial bee colony and fuzzy logic. They have focused on integrating the fitness function of an artificial bee colony with user comfort index and energy consumption.

Ali et al. in [68] used genetic programming and fuzzy logic for energy management. The model based on the comfort index of the single user means focusing on the complete house user comfort index and energy saving. The model was similar to their earlier model proposed in [66]. Another robust simulated model for energy consumption optimization was proposed by Ullah et al. in [69] using the genetic algorithm and particle swarm optimization. They have used sensors for the recording of the data and noise has been removed using a Kalman filter. The heating scenario has been considered in the study because the temperature in South Korea was already below the comfort index. The model for the comfort management and energy optimization in residential buildings based on fuzzy controllers and bat algorithm has been proposed by [70]. They have considered temperature, illumination and air quality as comfort parameters. The study has focused on South Korean environmental conditions.

### *4.2. Research Question 2: RQ2: Which algorithms/techniques, parameters, and pricing schemes have been used for energy consumption scheduling in smart homes?*

To answer the RQ2, we have extracted following relevant information from each article: 1) algorithm/technique, 2) focus area, 3) future energy prediction, 4) energy optimization, 5) thermal comfort, 6) visual comfort, 7) air quality comfort, 8) air quality, 9) temperature, 10) humidity, 11) illumination, 12) real-time pricing, 13) day ahead real-time pricing, 14) critical peak pricing, 15) time of use pricing, 16) flat pricing, 17) appliance power consumption based division, 18) optimal scheduling, 19) reduce peak to average ratio, 20) reduce power consumption, 21) reduce cost. We have provided detailed information of these parameters in Appendix C, Table A3.

4.2.1. Systematic Mapping of Scheduling Techniques

The data of Appendix C, Table A3, has been used for the systematic mapping same can be seen in Figure 3 [19]. The definition of the abbreviations used in the parameter portion of the map can be seen in the note portion of Appendix C, Table A3 and the remaining are provided as: earthworm optimization algorithm (EOA), genetic algorithm (GA), bacterial foraging optimization algorithm (BFOA), artificial fish swarm optimization (AFSO), mixed integer linear programming (MILP), daily maximum energy scheduling (DMES), sequential quadratic programming (SQP), crow search algorithm (CSA), cuckoo search optimization algorithm (CSOA), flower pollination algorithm (FPA), teacher learning based optimization (TLBO), dynamic programming algorithm (DPA), regression algorithm (RA), neural network (NN), support vector regression (SVR), random forest regression (RFR), distributed algorithm (DA), model predictive control (MPC), particle swarm optimization (PSO), temperature setpoint assignment (TSA), wind driven optimization (WDO), Jaya algorithm (JA), strawberry algorithm (SBA), and harmony search algorithm (HAS). In the scheduling portion total, 20 articles have been selected for the complete analysis the details are provided in Appendix C, Table A3.

We have observed that the researchers have not emphasized on comfort index in the scheduling techniques. Only 3 (15.00%) authors have considered the thermal comfort, and 1 (5.00%) visual comfort and temperature have been considered by 5 (25.00%). Regarding the pricing schemes, real-time pricing has been used by 9 (45.00%), critical peak pricing 8 (40.00%), a day ahead real-time pricing 4 (20.00%), time of use pricing 4 (20.00%) and flat pricing by 1 (5.00%).

The focus of all 20 (100%) articles was towards the energy saving and reducing power consumption. The second priority has been the optimal scheduling 19 (95.00%), followed by reducing the peak to average ratio of 17 (80.00%) and appliance power consumption-based division 15 (75.00%). Total 4 (20.00%) articles have considered the comfort index. Total 5 (25.00%) articles have focused the demand side management, and 1 (5.00%) author have considered the load management. Regarding the optimization algorithms for scheduling, the genetic algorithm has been used by 8 (40.00%) articles for the energy saving and cost reduction. For the comfort index, it has been used 2 (10.00%) times, for demand-side management 4 (20.00%) times, reduce the peak to average ratio 8 (40.00%) times, and for the load management 1 (5.00%) time. Total 3 (15.00%) authors have used bacterial foraging optimization algorithm for the energy saving, cost reduction and reduction of peak to average ratio. The other most used algorithm is the dynamic programming algorithm which has been used by 3 (15.00%) authors for the energy saving, cost reduction, reduction peak to average ratio, and 1 (5.00%) for the load management. The other most prominent algorithms remained neural networks, earthworm optimization algorithm, mixed integer linear programming, daily maximum energy scheduling, model predictive control, and particle swarm optimization.

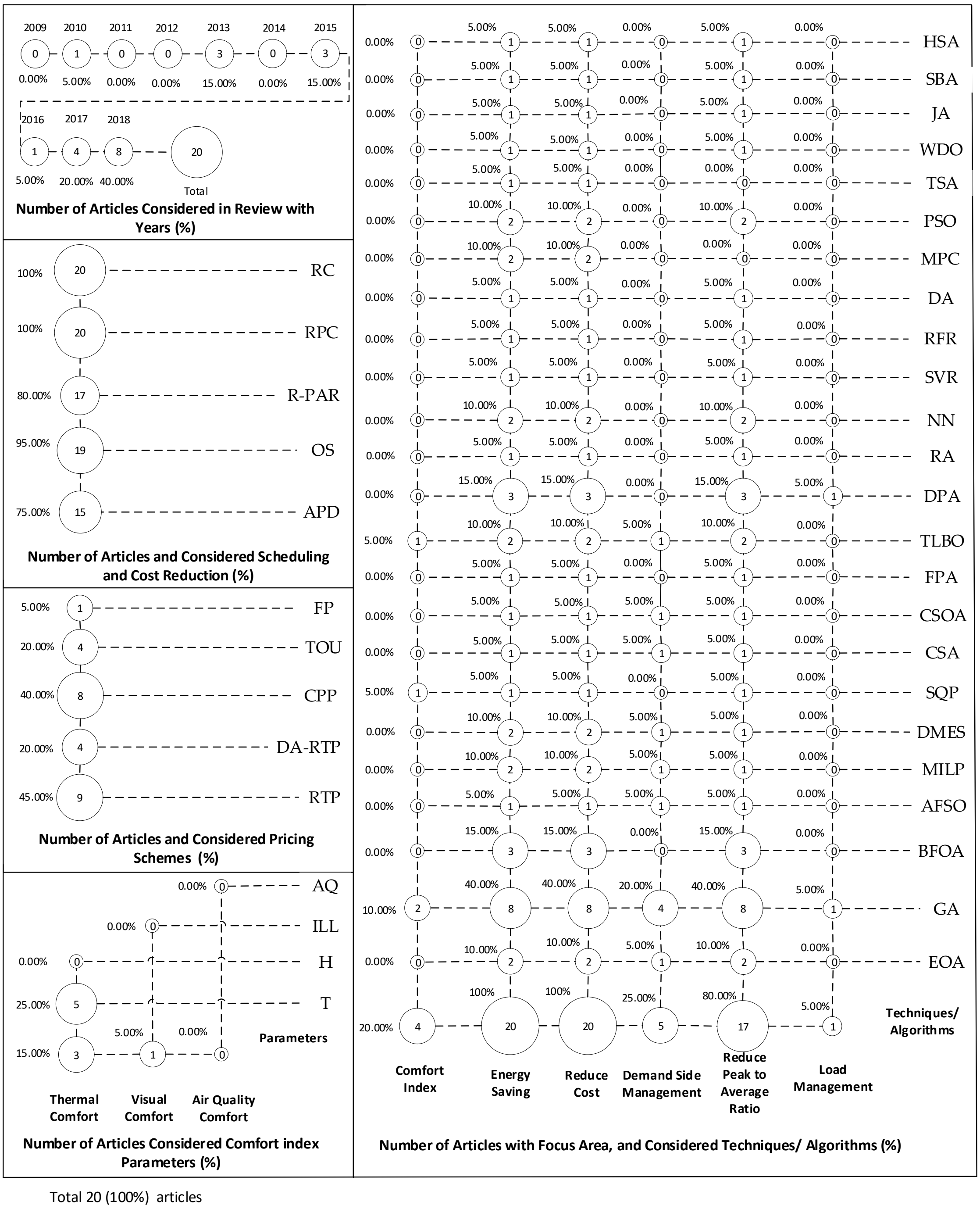

**Figure 3.** Visualization of a systematic map of scheduling techniques.

### 4.2.2. Algorithms / Techniques for Energy Optimization through Scheduling

The scheduling methods have benefits as well as drawbacks, like with the rescheduling of appliances from peak hours to regular hours. The cost of energy reduces, and consumers may reduce the electricity bills, but the user comfort also gets disturbed. In these methods, the users are restricted to use energy during low price hours. For the detailed comparative analysis kindly refer Appendix C Table A3.

Fuselli et al., in [71] have used neural network based action dependent heuristic dynamic programming (ADHDP) for the energy scheduling in the residential building using historical data. They have used particle swarm optimization (PSO) for the pre-training and enhancement of the optimization in the proposed algorithm. A model predictive methodology for the HVAC load management considers the inside temperature for the calculation of the cost of electricity [72]. The method provides energy efficient control to the customers for the energy management as per their requirement and preferences of inside temperature using temperature set-point assignment (TSA) algorithm. They have used the genetic algorithm for the optimized demand-side management, and the performance of GA based demand side management algorithm was better than the heuristic based evolutionary algorithms [73].

Ali et al. in [74], have presented a simulated model for the smart grid energy management and optimization using earthworm algorithm [75] and bacterial foraging algorithm [76,77]. Authors have used pricing scheme so that the consumers get an idea and use electricity accordingly and avoid usage of energy during peak hours to avoid the extra cost of electricity. Artificial fish swarm algorithm and genetic algorithm have been used to reduce the peak to average ratio (PAR) and cost of electricity by [78]. An electricity optimization rescheduling scheme using mixed integer linear programming method (MELP) and daily maximum energy scheduling device for South Africa has been proposed by [79]. Through the device, consumers can efficiently reduce electricity consumption.

Rasheed et al. in [80] have proposed an optimization technique to reduce the electricity bill and maintain the user comfort index. The home grid system contains a centralized energy management system (EMS), smart meter and home appliances, connected through a communication protocol. The proposed HVAC algorithm determines the time slot during which the electronic equipment operate with the minimum cost of electricity. A daily maximum energy scheduling for the smart homes that will work with the smart meters has been proposed by [81]. They have considered optimizing the daily energy consumption to reduce the cost of electricity bills while comfort index has not discussed. A service-oriented based modular multi-agent systems architecture for home energy management has been proposed by [82]. They have utilized visualization techniques and developed an online multiplayer game for real interaction with consumers for better rescheduling. They have developed a rescheduling algorithm for the cooling and heating plan.

Nadeem et al. in [83] proposed a rescheduling scheme for energy optimization focusing on the energy cost minimization keeping in mind the defined priorities of every appliance. They have used enhanced differential evolution (EDE) and teacher learning-based optimization (TLBO). The hierarchical model has been proposed in [84] which uses predictive control, for energy management. Khan et al., in [85] have used genetic and earthworm algorithms for energy management in smart grids, to minimize cost and Peak to average ratio (PAR) and maximize user comfort index. The smart grid has the physical system connected with the information system for monitoring activities of the system [14].

Mohsenian-Rad et al., in [86] have addressed the scheduling issue in smart meters with the deployed energy consumption scheduling (ECS) device. They have proposed a centralized power source shared among different consumers having ECS device having distributed algorithm installed in their smart meters. Javaid et al., in [87] used peak energy consumption to minimize the electricity cost and improve user comfort index using dynamic programming, genetic algorithm, and binary particle swarm optimization. For the communication of the appliances, they have used a home area network (HAN). Aslam et al., in [88] used genetic algorithm, cuckoo search optimization algorithm (CSOA) and crow search algorithm (CSA) for the reduction of electricity cost, waiting time and peak to average ratio (PART) in home energy management system (HEMS). For the electricity storage, they have incorporated smart electricity storage system (ESS) in the model. Awais et al., in [89] have used the bacterial foraging optimization algorithm (BFOA) and flower pollination algorithm (FPA) for the cost and PAR reduction in residential buildings.

Ahmad et al., in [90] integrated the renewable energy resources (RES) and energy storage system (ESS) to reduce the cost of electricity and PAR. To achieve the target, they have proposed an optimized home energy management system (OHEMS) using the day ahead pricing scheme. Samuel et al., in [91] focused on demand-side management in the residential sector using Earliglow based algorithm. The Earliglow algorithm uses the flavors of Jaya and strawberry algorithm (SBA). Hussain et al., in [92] proposed an efficient home energy management controller (EHEMC) to reduce energy cost and peak to average ratio (PAR) and maintain comfort index. The optimization has been achieved using genetic harmony search algorithm (GHSA). The comparison of the results of the algorithm with wind-driven optimization (WDO), harmony search algorithm (HSA), genetic algorithm (GA), has proved efficient technique for the scheduling, regarding cost reduction, PAR and comfort index. A demand-side management model for the residential load which helps to optimize the scheduling of power has been proposed by [93]. The day ahead pricing scheme has been considered for the scheduling. The teacher learning-based optimization has outperformed the other techniques regarding optimal scheduling.

### *4.3. Research Question 3: RQ3: How the edge and fog computing techniques are used in smart homes?*

To answer the RQ3, we have carried out a formal review of the edge and fog computing techniques. The edge and fog computing techniques for the optimization of energy and energy management and response time reduction is quite new domain therefore sufficient literature is not available. Hence most relevant studies have been reported here.

#### 4.3.1. Edge Computing, Fog Computing based Techniques in Smart Homes/Smart Grid

Ferrández-Pastor et al., in [94] proposed a method for smart devices based on edge computing, to tackle the issues related to scalability of services and interoperability. The model has automated the rule designing for better decision making based on installation behavior. They have used KNN and decision tree algorithms for energy generation and consumption management. They have used MQTT protocol as the communication protocol and Ubidots [95] IoT platform for experimentation. Froiz-Míguez et al., in [96] presented a ZiWi distributed fog computing home automation system (HAS) for the seamless communication among ZigBee and home devices connected through WiFi. They have used message queuing telemetry transport (MQTT) protocol for the communication between nodes of the smart home network. They have also discussed the home automation protocols and technologies, home automation systems for heterogeneous networks, fog computing architectures, and open source home automation systems. Further, they have carried out detailed discussion on comparison of the ESP8266 based boards.

Tehreem et al., in [97] have integrated the fog computing environment with the smart grid to manage the energy consumption efficiently. The region-based division has been carried out, having multigrid's assigned to each region. The balancing algorithms have been used to balance the load on virtual machines. They have used round robin (RR), throttled and beam search (BS) algorithms and the performance of BS algorithm was better as compared to the other algorithms. The model has three layers; cloud layer responsible for the communication between network and grid: the fog layer for the communication between cloud and microgrid, and the smart grid-based layer having multiple clusters for the communication with fogs. The fogs have been placed near to the end user to avoid latency rate. The BS algorithm has been proposed to achieve the minimum response time. Zakria et al., in [98] proposed a three-layered framework integrating fog computing to overcome the loading on cloud network in a smart grid environment. The focus of the model was to reduce the electricity consumers load from the cloud network. The connectivity with fog has been achieved using controllers. The second major layer contains fogs for receiving user requests and virtual machines. The cloud layer contains data centers and utility. They have used round robin (RR), throttled and shortest remaining time first (SRTF) approaches for the performance evaluation of the allocation of virtual machines. Further, they have also used closest data center service broker policy for the selection of fog. The SRTF can select the VM machines having a smaller number of requests. The fogs store customers data temporary before sending it to cloud for permanent storage. The performance of the proposed technique regarding cost was better than the other two techniques.

Zahoor et al., in [99] have proposed a fog-based model for the management of resources in a smart grid environment. The model has focused on hierarchical structures of the cloud-fog computing. They have used round robin, throttled and particle swarm optimization algorithm for the load balancing. The authors have focused on providing low latency services, and better scalability support for the grids. In the extended work, authors have used artificial bee colony (ABC) and ant colony optimization (ACO) in [100] for the energy management. Also, the authors have proposed the hybrid of ACO and ABC was known as hybrid artificial bee ant colony optimization (HABACO).

Fatima et al., in [101] proposed a fog computing based model for resource allocation in residential buildings. The performance has been measured using response time and processing time. The load balancing of virtual machines has been carried out using particle swarm optimization with simulated annealing (PSOSA). The broker policies include new dynamic service proximity, new dynamic response time and enhanced new response time. The simulations have been carried out using Cloud Analyst. Chekired et al., in [102] proposed a real-time dynamic pricing model for the charging and discharging of electric vehicles and energy management of residential buildings. The focus was to reduce the peak loads to save energy and reduce cost. The model use software defines networking (SDN), decentralized cloud architecture and network function virtualization (NFV). The model deals scheduling of the user requests in real time with supervised communication between the controllers of the microgrid. The other model focused on the electric vehicles for the scheduling and energy management using fog computing for the vehicle charging and discharging [103].

Butt et al., in [104] have proposed a technique for the improvement of efficiency of the smart grid in the cloud environment. Normally in the smart home network latency has been noticed due to a large number of requests from smart home clusters towards the smart grid which results in latency, delay and increased response time. The latency issue has been solved using fog computing to act as a middle layer between the cloud and smart grids. In the proposed methodology authors have used honey bee algorithm and compared results with round-robin algorithm regarding latency, delay and response time. The efficiency of the method has also been measured regarding the different broker policies like closet DC (CDC), closet DC (CDC), optimize response time (ORT), reconfigure dynamically with load (RDL) and advance service broker policy (ASP). Nazar et al., in [105] proposed architecture for the energy management system based on cloud and fog computing. The load management has been carried out using virtual machines on the fog servers using modified shortest job first (MSJF) algorithm. The results of the proposed technique were not better as compared to the round robin and throttled algorithms. They have deployed two patches of fog in the network connected to cloud and microgrid where the controller was responsible for maintaining the log of user requests. Ismail et al., [106] proposed fog architecture for the smart grid having a layered architecture. The virtual machines having the capability of multitasking has been used to address the multiple customer requests at the same time to reduce the response time. The allocation of the resources has been carried out using artificial bee colony algorithm. The performance of the artificial bee colony in this scenario was better than the round robin, particle swarm optimization and throttled algorithms.

Rehman et al., in [107] have proposed a four-layered architecture for the efficient resource management in cloud fog network. The technique aimed to reduce the load of customer requests from the cloud network. The balancing of the load has been carried out using round robin (RR), particle swarm optimization (PSO) and threshold-based load balancer (TBLB) algorithms. The TBLB has proved better results as compared to the other algorithms. Gao and Wu, in [108] presented a load balancing strategy in the cloud network using ant colony optimization algorithm (ACO). The focus of the scheme was to adjust the load and incoming requests toward the cloud dynamically. The candidate nodes have been found using forward-backward ant mechanism and max-min rules. The technique has been proved better regarding resource handling. Ashraf et al., in [109] have proposed a cloud-fog based model to avoid the wastage of energy in the smart grid environment. The resource requests have been handled using round robin, active monitoring virtual machine (AMVM), and throttled. The resource routing has been carried out using dynamic broker policy. Sharif et al., in [110] have proposed a model to deal with the maximum number of request in minimum time in a cloud network. The proposed insert select technique (IST) assists the data center in the optimized allocation of virtual machines to the requests. The simulation has been carried out in CloudSim.

Chakraborty and Datta in [111] proposed home automation based on edge computing, virtual IoT devices, and the internet of things technology. The home automation base of the edge computing technologies because the main functionalities of the smart homes depend on edge computing. The smart homes face the issue of the fragmentation due to the production of different types of data. The issue of fragmentation has been handled in the proposed method using a separate layer containing the drivers that help home devices to be part of the network. Lin and Hu, in [112] proposed a method for the residential energy consumption scheduling using particle swarm optimization algorithm. The method can be integrated with the edge computing.

Sun and Ansari, in [113] discussed the scalability of the internet of things architecture and proposed a mobile edge computing for the solution of scalability. The edgeIoT can handle data streams at the mobile edges. The computing resources are available at the fog nodes to reduce response time. Vallati et al., in [114] discussed the use of edge technology and its future in the smart home regarding the connectivity at the edges of the network. The future edge technology will allow the home appliances connectivity from different vendors in a single network for better communication. The protocols and standards for the home network are also the research topic of the researchers. The BETaaS EU project provides a runtime environment for the execution of applications in smart homes [115]. The researcher has used the standard CoAP proposed by [116]. The fog and edge computing technologies are quite new. The literature regarding these techniques in smart homes technologies is difficult to find. The focus of these techniques is to reduce the response time of the requests towards a smart grid, in this regard different researchers have considered the fog computing layers to efficiently manage the requests from smart home consumers towards the smart grids. The edge computing is the other growing field in the smart home technology which will shift the computing power towards the fog layers to avoid the delay and improve response time.

### *4.4. Research Question 4: RQ4: What are the technologies used in smart homes for the connectivity of devices?*

#### 4.4.1. Smart Homes

The smart homes are replacing traditional homes due to the advancement in technology. The smart homes contain intelligent devices, which are connected and can share information through a network. The intelligent devices are being used for the health monitoring of disable and older people [117]. In 2017 Gartner, firm defined connected home solutions as *"Connected home solutions consist of a set of devices and services that are connected and the internet and can automatically respond to preset rules, be remotely accessed and managed by mobile apps or a browser, and send alerts or messages to the user(s)"* [118].

The smart homes are automated and digitized and provide enhanced monitoring, comfort, energy conservation, maintenance, home activities and security of occupants. The residents can access its home functions through a display and controllers, like built-in display, computers, tablets or mobile phones [117]. The functionality of smart homes differs from complete building to a home; it depends upon the requirements of the resident. The smart home is a combination of automation digitization and interconnectivity of home automation areas, like room control, light control, security control and so forth. The internet of things has made it possible to connect the smart meters, smart televisions, entertainment systems, security systems and so forth. The field of smart homes is progressing, and full-fledged smart home is not a reality today. Like other areas of smart homes, there is a need for the improvement of HVAC systems regarding energy efficiency and improvement in user comfort [117]. According to Jhonson Controls Energy Efficiency survey 2017, *"70 percent of organizations are paying more attention to energy efficiency than the year before and that 58 percent expect an increase of energy efficiency investments in 2018"* [119]. The need for energy efficiency will grow further with the advancement of IoT and smart technologies in the buildings. "*Energy management in smart homes mainly comprises of smart meters, smart appliances, energy management system for devices, and home power generation, all functioning on smart grids*" [119].

The advancements in smart homes are adding more comfort in living styles, more convenience, a bundle of entertainment and sustainability [117]. In the past, due to technological prices of devices, the smart homes were expensive but now gradually becoming in reach of everyone. The intelligent assistant utilizes artificial intelligence techniques so that user can communicate with the devices through voice control. The devices can communicate through the advanced wireless networks; data can be saved on the cloud, technologies becoming more energy efficient and opening an era for smart homes. Smart homes are now aiming to shift towards smart space, to solve personal and social issues [117]. The use of edge and fog computing is the other area of smart homes advancement regarding reduction in response time and delay.

### *4.4.2. Technologies Used in Smart Homes*

Researchers are trying to solve the complex issues for the proper implementation of smart home systems so that the users can get maximum benefits. In smart homes network, the devices can communicate with each other through specific protocols like Zigbee, KNX, and Z wave. Zigbee is a wireless network it uses a device to relay a signal to another device, strengthening and expanding the network. Zigbee can be built in dimmers, door locks, thermostats, and many more. Z-Wave is a mesh network using low-energy radio waves that provides wireless connectivity to the home devices [120,121]. Z wave is like Zigbee but slower in performance [120,121]. KNX is open source and mostly used for the automation, and each system in KNX is smart itself, the most prominent advantage is if a system fails it does not affect the performance of other devices connected to the network. It operates on more than one physical layers [120,121]. The one-way communication protocol is mostly used in smart homes [120,121]. The second choice of protocol is Ethernet which is much faster than wireless networks in which the smart home devices are get connected through the wires. The other network is Wi-Fi which is more convenient and can be used up to a range of 25 meters for the connectivity of devices. For the shorter-range communications, Bluetooth is preferred in the smart homes [120,121]. In the smart home's network the energy management controller control the electronic devices [74]. The electronic devices consume considerable energy depending on their nature like every smart home has an air conditioner for the Colling, a heater for the heating, purifiers for the air quality and light for the illumination. We can say that mostly these three types of equipment are used in the houses for the comfort of residents [122].

### *4.4.3. Optimization in Domains of IoT based Smart Cities*

We cannot deny the smart cities by keeping in view the current progress in technology; in the next ten to twenty years there will be considerable advancement in smart cities. The developing countries are already moving toward the smart cities [123,124]. There are different domains of IoT where optimization can be applied if we narrow it down to the smart homes then it can be used for the surveillance, comfort, and automation. If we consider it in point of view of energy management then can be used in smart metering and smart grids. However, regarding the optimization still, a lot needs to be done due to the challenges regarding optimization algorithms, and proper selection of parameters like temperature, air quality, humidity, clothing, lighting and so forth. The focus of this paper is on smart homes and a particular in comfort, automation and a minor part on the energy management side as well as shown in Figure 4 [125].

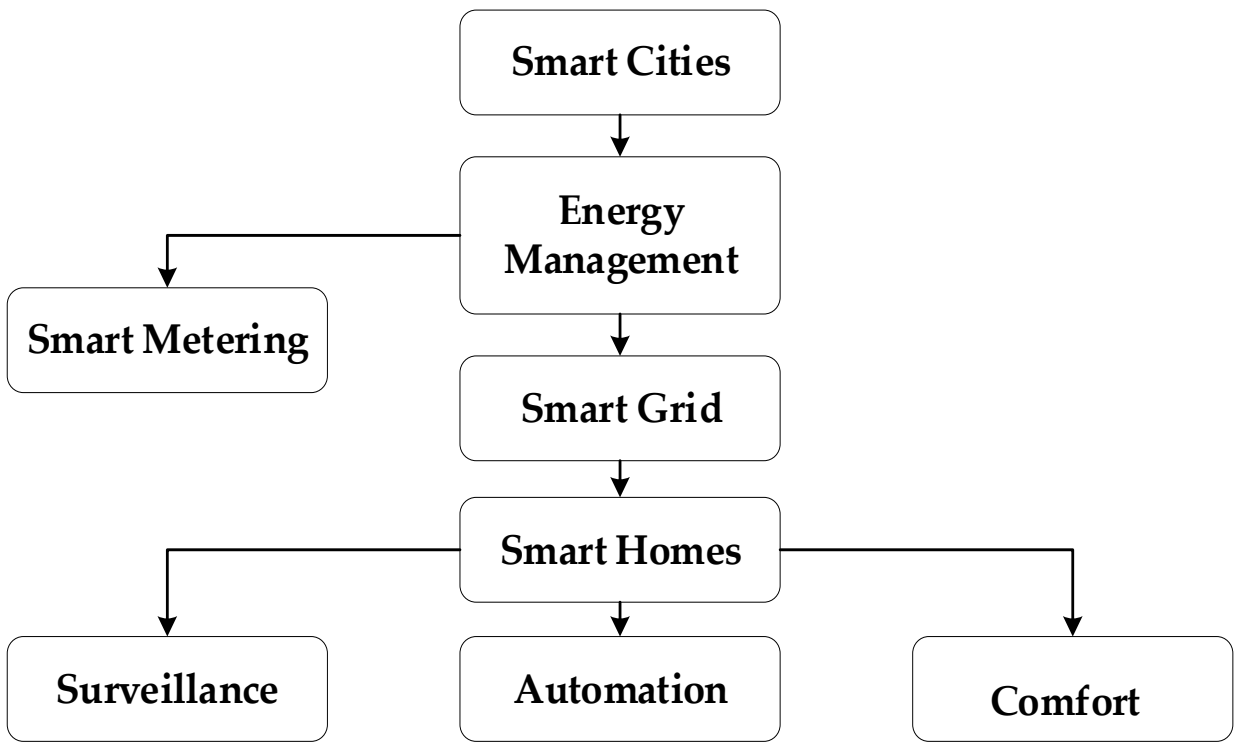

**Figure 4.** Role of optimization in smart homes

### *4.5. Research Question 5: RQ5: What are the different comfort index parameters in smart homes?*

The comfort in smart homes is a certain range of parameters like temperature, humidity, illumination, air quality where the residents feel comfortable and their productivity increase. The focus of this study is the thermal comfort, visual comfort, and air quality comfort.

### 4.5.1. Thermal Comfort

According to Fanger [126,127], the main contributors towards the thermal comfort are as shown in Figure 5.

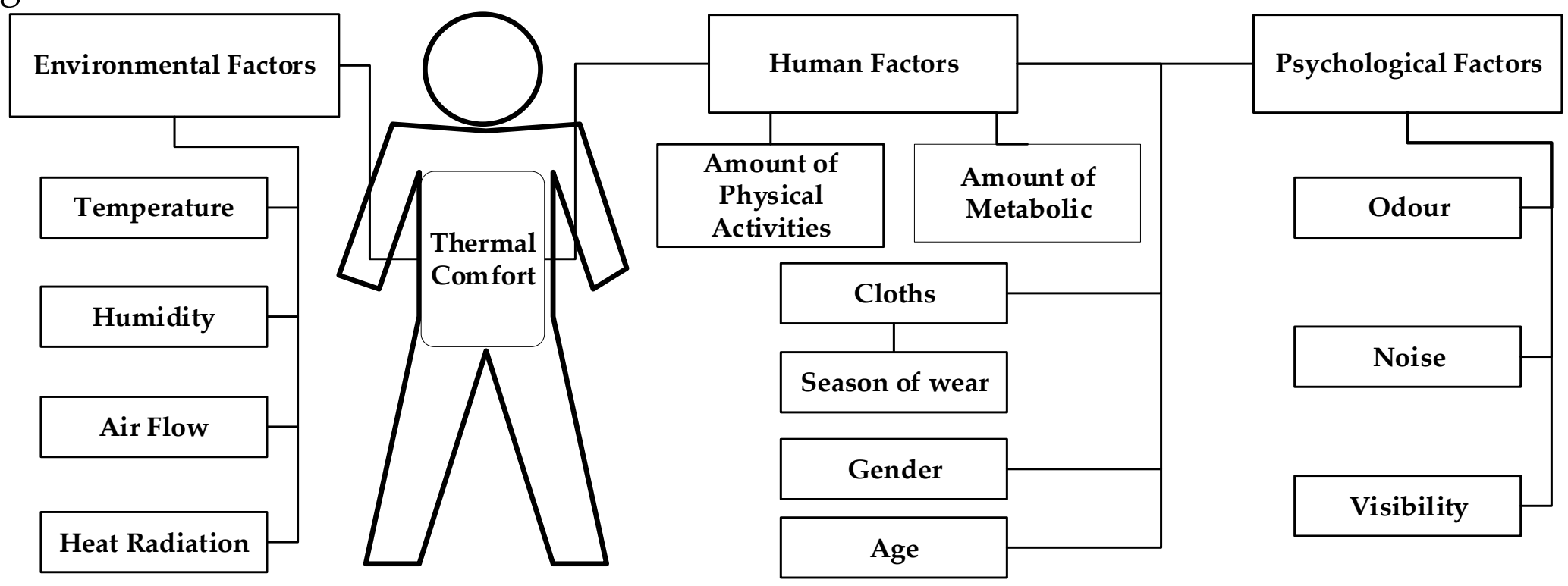

**Figure 5.** Categorization of different thermal comfort factors [126,127].

The noise regarding the home will be the noise of electronic pieces of equipment, like fans, AC, music system and so forth [128]. Most of the authors have considered the temperature as thermal comfort, but as per Malaysian environment, the consideration of temperature, along with humidity, for the thermal comfort is necessary for the smart homes [128]. The environmental factors are the combination of temperature, humidity, air flow, and heat radiation. In the current scenario; the environmental factors mean the internal environment inside the smart home [128]. On the other side, human factors include; the number of physical activities, amount of metabolic, besides that other factors are the clothes, gender, age, and season of wear. The psychological factors also affect the comfort index which includes odor, noise, and visibility [128].

### 4.5.2. Visual Comfort

One of the main factors of the overall comfort index is visual comfort. The visual comfort is very important in the areas like study room in the home, office building and so forth. The components of the visual comfort can be seen in Figure 6; the same has been retrieved from references [129,130].

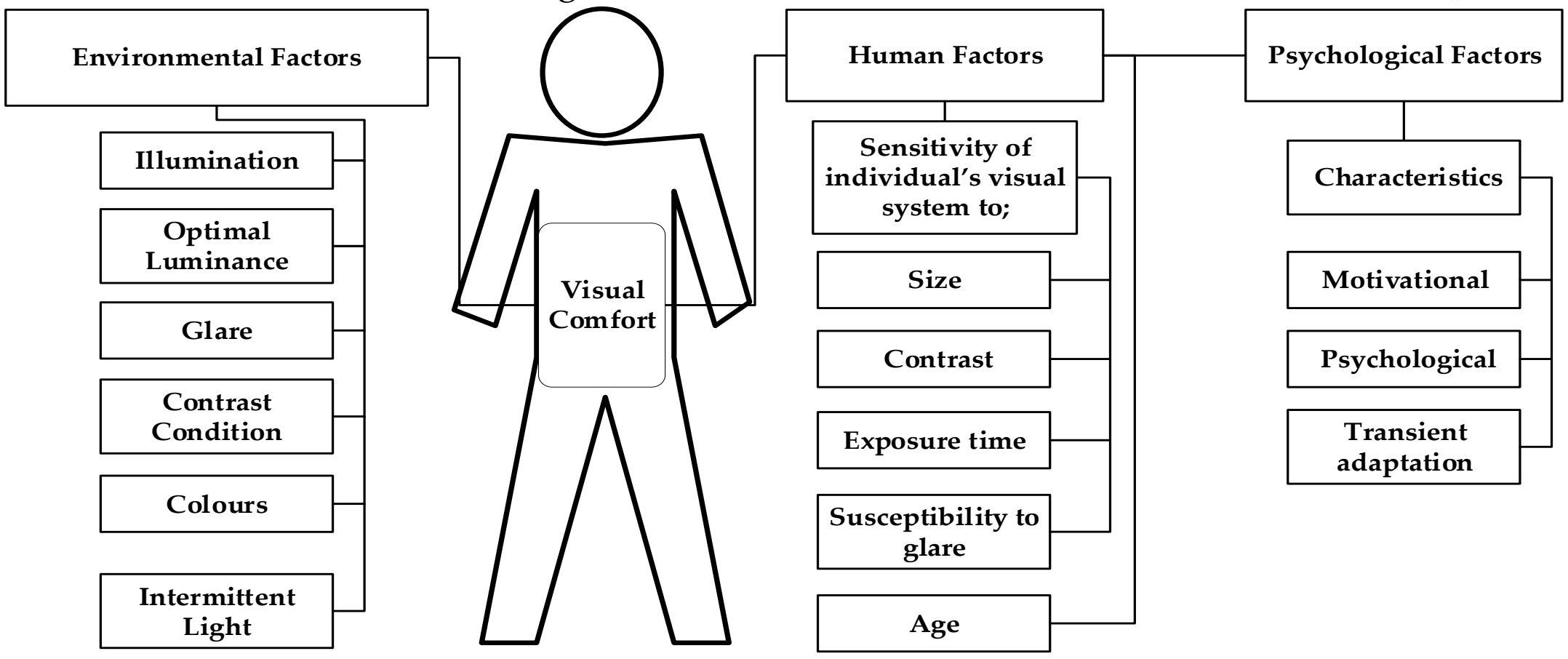

**Figure 6.** Categorization of different visual comfort factors [129,130].

The environmental factors that determine the visual comfort are uniform illumination, optimal luminance, no glare, sufficient contrast conditions, correct colors, and absence of stroboscopic effects or intermittent light [129,130]. The human factors include the sensitivity of an individual's visual system to size, contrast, exposure time, susceptibility to glare. The other main human factor is the age, because with the increase in age the eyesight decreases which cause problems. Mostly the elderly persons require lighter as compared to the younger persons. The psychological factors include the motivational, psychological and transient adaptation characterization. The measurement of glare and other conditions is the most difficult task; therefore, the data of glare is not available, and most of the researchers have considered the illumination as visual comfort measurement. The factors that determine the visual comfort are uniform illumination, optimal luminance, no glare, adequate contrast conditions, correct colors, and absence of stroboscopic effects or intermittent light.

#### 4.5.3. Air Quality Comfort

The maintenance of air quality comfort is the most critical component of smart homes. The desired concentration of $CO_2$ is 800ppm [131] if the range of $CO_2$ increases to this level it may cause serious issues and can cause death as well. In traditional homes, the amount of $CO_2$ can be balanced with the opening and closing of the windows, but with the introduction of smart homes, the devices like air purifiers are available in the market which helps to keep the air quality to comfort zone. Mostly the air quality is measured regarding $CO_2$; the other factors include a total volatile organic compound and volatile organic compound in the air [131]. The different factor of air quality can be seen in Figure 7.

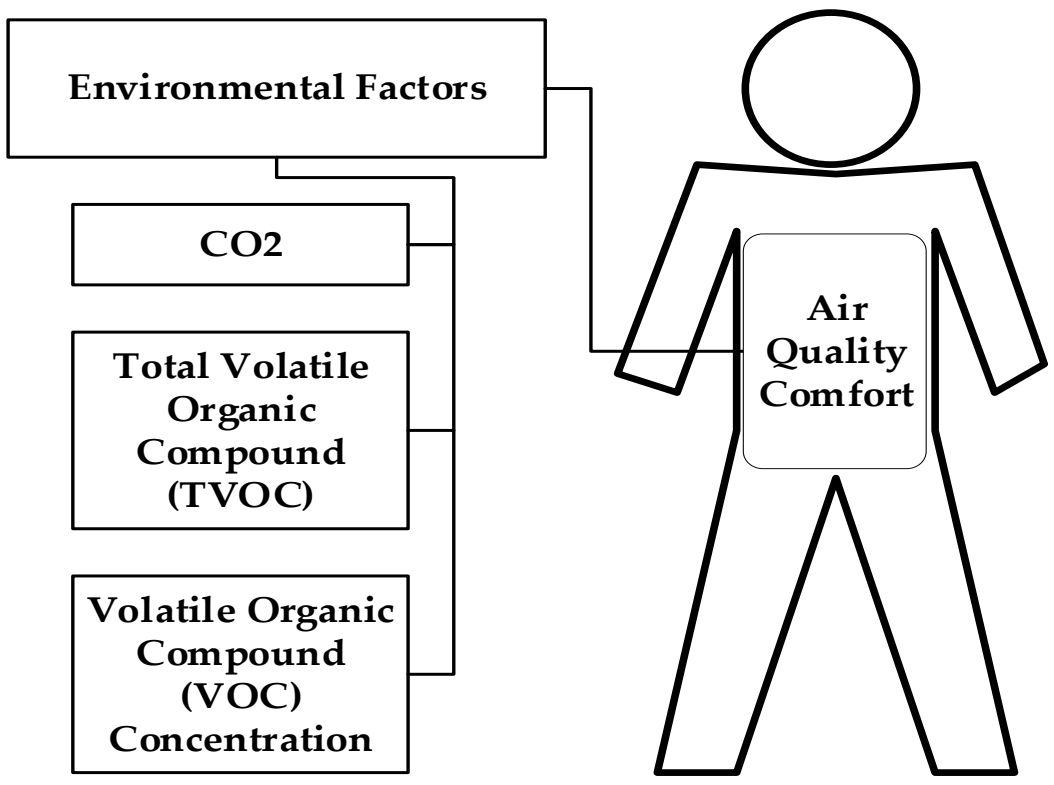

**Figure 7.** Categorization of different air quality comfort factors [131].

### *4.6. Limitations of the study:*

The review relied on a relatively limited number of databases for the identification of potentially eligible studies. Limited search strings have been used to identify the literature. We have not included the literature published before 1996 in the study.

## 5. Conclusion

The researchers have considered every factor to satisfy the comfort index and energy consumption reduction in the residential building using different optimization algorithms; it can be concluded that genetic algorithms have performed better as compared to the other algorithms. The most prominent gaps can be seen regarding the consideration of parameters and preference of the users. Mostly the researchers have not considered the humidity in their experimentation for the comfort index calculation. The reason for not considering humidity might be the environmental conditions of the countries where research has been conducted. In the ecological conditions of Malaysia humidity is the critical factor because the humidity remains outside of the comfort zone; therefore, we cannot ignore the consideration of humidity. Majority of the researchers have not considered the multiple user comfort inside buildings. Mostly the focus of studies was to satisfy the inside environment according to the standards set for the building environment. Further very few researchers have considered the external environment conditions in their experimentation which need attention to fill the research gap. Very few researchers have considered air flow (AF), heat radiation/flow (HR/F), cloth thermal insulation (CTI), metabolic rate (MR), and water vapor pressure (WVP).

Further, it has been identified that researchers had introduced some new optimization algorithms that can be used for the optimization of energy consumption in smart homes. Although some of them have already been used for the scheduling problem in smart grid and homes. These algorithms includes; earthworm optimization algorithm [75], Firefly [132], bat algorithm [133], anarchic society optimization (ASO) algorithm [134], cuckoo optimization algorithm (COA) [135], league championship algorithm (LCA) [136,137], and crow search algorithm (CSA) [138]. The original firefly algorithm faces some drawbacks due to which the variations of the firefly algorithm were proposed and used for the energy optimization and various other optimization problems in different areas. The researchers are now focusing on firefly optimization algorithm for the optimization of energy consumption [139]. The crow search algorithm is the newly proposed optimization algorithm by Askar Zadeh in 2016 [140]. The algorithm has been used for energy optimization by decidedly fewer researchers, the natural process behind the crow search algorithm has been explained by [138,141]. The CSA focusses on the two parameters flight length and awareness probability. Therefore the algorithm is suitable to be used for optimization problems like energy optimization [142].

## 6. Future Work Directions

The automatic prediction of user parameters will ultimately improve the comfort index and accommodate more users of a residential building. The automatic user parameters will also help the children and person with disabilities to operate the system with ease. Currently, the membership function selection is manual, and once the parameters are selected at the initial level, the user cannot change. The automatic selection of membership function parameters can be the research area to predict the parameters for fuzzy logic membership functions which can dynamically change according to the weather conditions. Regarding the scheduling techniques no doubt they provide flexibility to operate the appliances in off-peak hours to reduce the energy consumption, but mostly the user comfort is compromised. We have observed that the comfort index in scheduling techniques have been measured regarding waiting time instead of the comfort indexes in the case of optimization techniques. Where in the optimization techniques most of the researchers have considered the thermal comfort, visual comfort, and air quality comfort. Furthermore, as per observation, the scheduling techniques are mostly focused toward the smart grid instead of smart homes. These techniques can be useful for the scheduling in smart grids. The optimization technique provides both comfort index and energy consumption optimization at the same time. The fuzzy controller is the best choice in optimization techniques. Fuzzy controllers are more comfortable to design as compared to other controllers, but they require expertise as well. The fuzzy controllers have now replaced the PID controllers most of the device industries are using fuzzy controllers. The PID controllers design is analytical while the fuzzy controller's design is intuitive. The fuzzy controllers are real-time expert systems implementing human experience and knowledge. With limited knowledge of the system, we can achieve much better automation as compared to the other controllers. On the other side, the success of fuzzy controllers depends on human expertise, without human expertise we cannot achieve better results.

The network-based technologies are still in improving phase, as per observations it has been noticed that the proper communication protocol for the smart homes is not present which leads towards the security issues in smart homes. Further, the flexibility is also not available in smart homes regarding the appliances from different vendors. The users are bound to purchase all the home appliances from the same vendor to avoid the connectivity problem although the research is in progress to create a general protocol which can connect devices from different vendors without any communication issue. The researchers have applied the fog computing techniques incorporated in the cloud network to reduce the delay and response time. In this regard, many architectures have been proposed to provide a layer of fog between the cloud and smart grid to reduce latency and deal with the customer requests efficiently. The edge computing is another progressing field in the smart homes and IoT networks which provided computing power to the fog and edges of the cloud network to avoid the delay.

Authors have proposed methods for energy consumption optimization, and prediction, without considering the multi-user comfort index. The idea of the hybrid method for the energy consumption optimization and maximization of user comfort index using different optimization algorithms will cover the major limitations of already proposed techniques. The consideration of multi-user comfort index is as important as the optimization of energy consumption. The demand of every user is different like a person feels too much coldness on 24 degrees, but the other user of the house feels comfortable with 24 degrees. So, the future work of this study is to satisfy every resident of the house by selecting the best range of comfort parameters and considering the minimum possible energy consumption. This problem can be solved with the introduction of an entirely new technique or the hybrid of already proposed methods so that both factors of energy consumption minimization and user comfort maximization simultaneously be achieved. The consideration of multiple user comfort index will benefit more residents as compared to the single user models as elaborated in Figure 8.

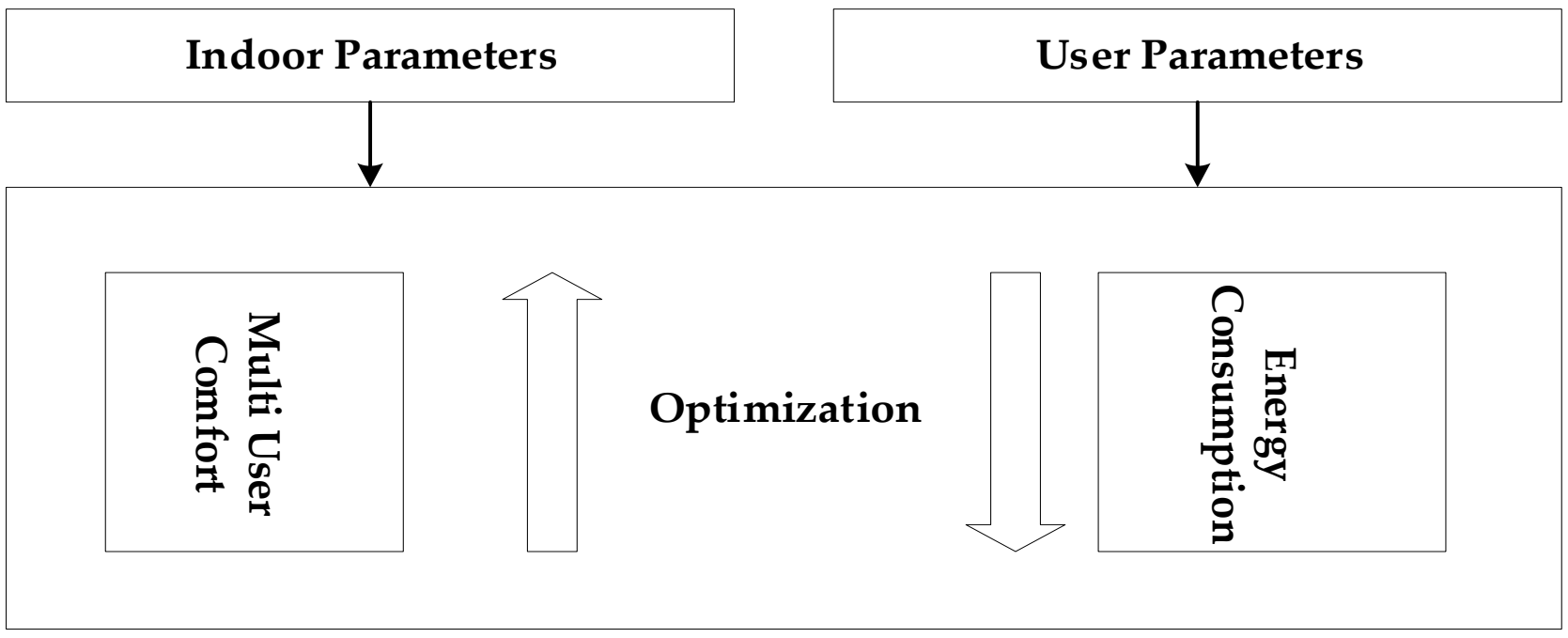

**Figure 8.** The objective of future study in smart homes

**Author Contributions:** "Conceptualization, A.S.S. and H.N.; Methodology, A.S.S., H.N., and M.F.; Investigation, A.S.S., Validation, A.S.S., H.N., M.F., and A.L.; Formal Analysis, M.F.; Resources, H.N.; Data Curation, A.S.S.; Writing-Original Draft Preparation, A.S.S..; Writing-Review & Editing, A.S.S. and H.N.; Visualization, A.S.S.; Supervision, H.N., and A.L.; Project Administration, H.N. and A.L; Funding Acquisition, H.N. Forward- and backward snowballing technique, A.S."

**Funding:** "This research received no external funding."

**Acknowledgments:** The authors gratefully acknowledge the assistance of the people, who reviewed the manuscript, and Dr. Kamran Khowaja, Isra University, Pakistan, for his valuable suggestions regarding the restructuring of the paper and systematic mapping. Any correspondence related to this paper should be addressed to Dr. Haidawati Nasir.

**Conflicts of Interest:** The authors declare no conflict of interest.

## Appendix A

**Table A1.** Articles selected for the review

| Sr. No. | Reference | Sr. No. | Reference | Sr. No. | Reference | Sr. No. | Reference |
|---|---|---|---|---|---|---|---|
| 1. | Li et al., [5] | 2. | Esmat et al., [11] | 3. | Butt et al., [12] | 4. | Amin et al., [13] |
| 5. | Zhao et al., [14] | 6. | Wang et al., [17] | 7. | Dounis et a;., [20] | 8. | Huang and Lam [21] |
| 9. | Wang et al., [22] | 10. | Kolokotsa et al., [23] | 11. | Morel et al., [24] | 12. | Wright et al., [25] |
| 13. | Kolokotsa et al., [26] | 14. | Kolokotsa et al., [27] | 15. | Moon et al., [28] | 16. | Calvino et al., [29] |
| 17. | Fong et al., [30] | 18. | Trobec Lah et al., [31] | 19. | Doukas et al., [32] | 20. | Dalamagkidis et al., [33] |
| 21. | Liang et al., [34] | 22. | Fountain et al., [35] | 23. | Freire et al., [36] | 24. | Chi-Min et al., [37] |
| 25. | Mitsios et al., [38] | 26. | Moon et al., [39] | 27. | Navale, et al., [40] | 28. | Dounis et al., [41] |
| 29. | Klein et al., [42] | 30. | Khan et al., [43] | 31. | Ghahramani et al., [44] | 32. | Nassif et al., [45] |
| 33. | Scherer et al., [46] | 34. | Mousavi et al., [47] | 35. | Carlucci et al., [48] | 36. | Nagy et al., [49] |
| 37. | Chou et al., [50] | 38. | Chew et al., [51] | 39. | Galbazar et al., [52] | 40. | Delgarm et al., [53] |
| 41. | Shaikh et al., [54] | 42. | Shaikh et al., [55] | 43. | Zheng et al., [56] | 44. | Lim et al., [57] |
| 45. | Park et al., [58] | 46. | Xu [59] | 47. | Putra [60] | 48. | Ain et al., [61] |
| 49. | Marvugila et al., [62] | 50. | Collotta et al., [65] | 51. | Ali et al., [66] | 52. | Wahid et al., [67] |
| 53. | Ali et al., [68] | 54. | Ullah et al., [69] | 55. | Fayaz et al., [70] | 56. | Fuselli et al., [71] |
| 57. | Avci et al., [72] | 58. | Bharathi et al., [73] | 59. | Ali et al., [74] | 60. | Talha et al., [78] |
| 61. | Longe et al., [79] | 62. | Rasheed et al., [80] | 63. | Longe et al., [81] | 64. | Akasiadis et al., [82] |
| 65. | Javaid et al., [83] | 66. | Lefort et al., [84] | 67. | Sajawal ur Rehman et al., [85] | 68. | Mohsenian-Rad et al., [86] |
| 69. | Javaid et al., [87] | 70. | Aslam et al., [88] | 71. | Awais et al., [89] | 72. | Ahmad et al., [90] |
| 73. | Samuel et al., [91] | 74. | Hussain et al., [92] | 75. | Manzoor et al., [93] | 76. | Ferrández-Pastor et al., [94] |
| 77. | Froiz-Míguez et al., [96] | 78. | Tehreem et al., [97] | 79. | Zakria et al., [98] | 80. | Zahoor et al., [99] |
| 81. | Zahor et al., [100] | 82. | Fatima et al., [101] | 83. | Chekired et al., [102] | 84. | Chekired et al., [103] |
| 85. | Butt et al., [104] | 86. | Nazar et al., [105] | 87. | Ismail et al., [106] | 88. | Rehman et al., [107] |
| 89. | Gao and Wu, [108] | 90. | Ashraf et al., [109] | 91. | Sharif et al., [110] | 92. | Chakraborty and Datta in [111] |
| 93. | Lin and Hu, [112] | 94. | Sun and Ansari, [113] | 95. | Vallati et al., [114] | 96. | Vallati et al., [115] |
| 97. | Osipov [120] | 98. | Bharathi [121] | 99. | Risteska et al., [122] | 100. | Ejaz et al., [125] |
| 101. | Fanger [126] | 102. | Fanger [127] | 103. | Busl [128] | 104. | Silvester and Konstantinou [129] |
| 105. | Rajendrakumar [130] | 106. | Batterman [131] | 107. | Yang [132] | 108. | Yang and Hossein Gandomi [133] |
| 109. | Bozorgi et al., [134] | 110. | Jafari et al., [135] | 111. | Husseinzadeh Kashan [136] | 112. | Rezaei et al., [137] |
| 113. | Zolghadr-Asli et al., [138] | 114. | Wahid et al., [139] | 115. | Askar Zadeh in 2016 [140] | | |

## Appendix B

**Table A2.** Energy optimization techniques and their comparative analysis

| Ref. | Algorithm / Techniques | Focus Area | Optimization/ Prediction | | Comfort Index Parameters | | | | | | | | | | | | | | | | Number of Users/ Rooms | | |
|---|---|---|---|---|---|---|---|---|---|---|---|---|---|---|---|---|---|---|---|---|---|---|---|
| | | | EP | EO | TC | VC | AQC | SC | AQ | T | H | Ill | AF | HR/F | CTI | MR | WVP | EE | SU | MU | IR |
| [17] | Particle Swarm Optimization (PSO), Hierarchical multi-agent theory, Fuzzy Controller. | Energy Saving, Comfort Index. | × | ✓ | ✓ | ✓ | ✓ | × | ✓ | ✓ | × | ✓ | × | × | × | × | × | × | ✓ | × | × |
| [22] | Genetic Algorithm, PID Controller. | Thermal Comfort. | × | ✓ | ✓ | × | ✓ | × | ✓ | ✓ | ✓ | × | ✓ | × | × | × | × | ✓ | ✓ | × | × |
| [23] | Fuzzy PD Controller. | Comfort Index | × | × | ✓ | ✓ | ✓ | × | ✓ | ✓ | × | ✓ | × | × | × | × | × | ✓ | ✓ | × | × |
| [24] | Artificial Neural Network. | Thermal Comfort | ✓ | ✓ | ✓ | × | × | × | × | ✓ | × | × | × | × | × | × | × | ✓ | ✓ | × | ✓ |

| | | | | | | | | | | | | | | | | | | | | |
|---|---|---|---|---|---|---|---|---|---|---|---|---|---|---|---|---|---|---|---|---|
| **[25]** | Multi-Objective Genetic Algorithm (MOGA). | Thermal Comfort | × | × | ✓ | × | × | × | × | ✓ | × | × | ✓ | × | × | ✓ | × | × | ✓ | × | × |
| **[26]** | Fuzzy Logic, Genetic Algorithm. | Comfort Index | × | ✓ | ✓ | ✓ | ✓ | × | ✓ | ✓ | ✓ | ✓ | × | × | × | × | × | ✓ | ✓ | × | × |
| **[27]** | Fuzzy P, Fuzzy PID, Fuzzy PI, Fuzzy PD, Adaptive Fuzzy PD. | Comfort index | × | ✓ | ✓ | ✓ | ✓ | × | ✓ | ✓ | × | ✓ | × | × | × | × | × | ✓ | ✓ | × | × |
| **[29]** | Fuzzy Adaptive PID Controller. | Thermal Comfort | × | × | ✓ | × | × | × | × | × | × | × | × | ✓ | ✓ | ✓ | ✓ | ✓ | ✓ | × | × |
| **[31]** | Fuzzy Logic, Auxiliary PID Controller, IDR BLOCK. | Visual Comfort, Energy Saving | × | ✓ | × | ✓ | × | × | × | × | × | ✓ | × | × | × | × | × | × | × | × | × |
| **[32]** | Decision Support Model. | Comfort Index, Energy Saving, Rule Base | × | ✓ | ✓ | ✓ | ✓ | × | ✓ | ✓ | ✓ | ✓ | × | × | × | × | × | ✓ | ✓ | × | ✓ |
| **[33]** | Fuzzy PD Controller, Linear Reinforcement Learning Controller (LRLC). | Comfort Index | ✓ | ✓ | ✓ | ✓ | ✓ | × | ✓ | ✓ | × | × | × | × | × | × | × | × | ✓ | × | × |
| **[34]** | Predicted Mean Vote (PMV), Human Learning, Direct Neural Network Controller. | Comfort Index, Energy Saving | × | ✓ | ✓ | × | × | × | × | ✓ | ✓ | × | ✓ | × | × | ✓ | × | ✓ | ✓ | × | × |
| **[36]** | Model-Based Predictive Control Strategy. | Thermal Comfort, Energy Saving | × | ✓ | ✓ | × | × | × | × | ✓ | ✓ | × | × | × | ✓ | ✓ | × | × | ✓ | × | × |
| **[38]** | Developed new Control Algorithm. | Comfort Index, Energy Saving | × | ✓ | ✓ | ✓ | ✓ | × | ✓ | ✓ | × | ✓ | × | × | × | × | × | ✓ | ✓ | × | × |
| **[39]** | Artificial Neural Network, Predicted Mean Vote (PMV). | Thermal Comfort, Energy Saving. | × | ✓ | ✓ | × | × | × | × | ✓ | ✓ | × | ✓ | × | × | × | × | ✓ | ✓ | × | × |
| **[41]** | Fuzzy Logic Controller, Genetic Algorithm. | Comfort Index, Energy Saving, Learning to Control. | × | ✓ | ✓ | ✓ | ✓ | × | ✓ | ✓ | × | ✓ | × | × | × | × | × | × | ✓ | × | × |
| **[42]** | Markov Decision Problems (MDP). | Thermal Comfort, Energy Saving. | × | ✓ | ✓ | × | × | × | × | ✓ | × | × | × | × | × | × | × | ✓ | × | × | × |
| **[43]** | Genetic Algorithm, Fuzzy Logic Controller. | Comfort Index, Thermal Comfort, Energy Saving. | × | ✓ | ✓ | × | × | × | × | ✓ | ✓ | × | × | × | × | × | × | × | ✓ | × | × |
| **[44]** | Knowledgebase, Heuristic System Identification Approach, Fuzzy Pattern Recognition, Spearman's Rank Correlation Analysis. | Thermal Comfort, Energy Saving. | ✓ | ✓ | ✓ | × | × | × | × | ✓ | ✓ | × | ✓ | × | × | × | × | × | × | ✓ | ✓ |
| **[45]** | Genetic Algorithm, Artificial Neural Network. | Comfort Index, | × | ✓ | ✓ | × | × | × | × | ✓ | ✓ | × | ✓ | × | × | × | × | × | × | × | × |

| | | | | | | | | | | | | | | | | | | | | |
|---|---|---|---|---|---|---|---|---|---|---|---|---|---|---|---|---|---|---|---|---|
| | | Energy Saving. | | | | | | | | | | | | | | | | | | |
| **[46]** | Distributed Model Predictive Control (DMPC). | Energy Saving. | × | ✓ | ✓ | × | × | × | × | ✓ | × | × | × | × | × | × | × | × | ✓ | × | × |
| **[49]** | Passive Infrared (PIR) Motion sensors. | Visual Comfort, Energy Saving. | × | ✓ | × | ✓ | × | × | × | × | × | ✓ | × | × | × | × | × | × | × | × | ✓ |
| **[51]** | Passive Infrared (PIR) Sensor, TEMT6000 Ambient Light Sensor, Daylight Harvesting, ZigBee. | Visual Comfort, Energy Saving. | × | ✓ | × | ✓ | × | × | × | × | × | ✓ | × | × | × | × | × | × | × | × | ✓ |
| **[52]** | Ant Colony Optimization Algorithm, Fuzzy Controller. | Comfort Index, Energy Saving. | × | ✓ | ✓ | ✓ | ✓ | × | ✓ | ✓ | × | ✓ | × | × | × | × | × | × | ✓ | × | × |
| **[53]** | Mono- and Multi-Objective Particle Swarm Optimization (MOPSO), Weighted Sum Method (WSM). | Energy Saving. | × | ✓ | ✓ | ✓ | × | × | × | ✓ | × | ✓ | × | × | × | × | × | × | ✓ | × | ✓ |
| **[55]** | Stochastic Intelligent Optimization, Multi-Objective Genetic Algorithm (MOGA), Hybrid Multi-Objective Genetic Algorithm (HMOGA), Fuzzy Logic. | Comfort Index, Energy Saving. | × | ✓ | ✓ | ✓ | ✓ | × | ✓ | ✓ | × | ✓ | × | × | × | × | × | × | ✓ | × | ✓ |
| **[57]** | EnergyPlus Runtime Language (Erl). | Thermal Comfort, Occupant Behavior, Energy Saving. | ✓ | ✓ | ✓ | × | × | × | × | ✓ | × | × | × | × | × | × | × | ✓ | × | × | × |
| **[59]** | SoftMax Regression, Multinomial Logistic Regression, Building Control Virtual Test Bed (BCVTB). | Multiple occupant’s comforts, Energy Saving. | × | ✓ | ✓ | × | × | × | × | ✓ | × | × | × | × | × | × | × | × | × | ✓ | × |
| **[61]** | Fuzzy Logic, Mamdani and Sugeno Fuzzy Inference Systems. | Thermal Comfort, Energy Saving | × | ✓ | ✓ | × | × | × | × | ✓ | ✓ | × | × | × | × | × | × | ✓ | ✓ | × | × |
| **[62]** | Auto-Regressive Neural Network with External Inputs (NNARX), Fuzzy Logic. | Thermal Comfort | ✓ | ✓ | ✓ | × | × | × | × | ✓ | ✓ | × | ✓ | × | × | × | × | × | ✓ | × | × |
| **[65]** | Neural Network, Fuzzy Controller. | Thermal Comfort, Energy Saving. | × | ✓ | ✓ | × | × | × | × | ✓ | × | × | × | × | × | × | × | × | ✓ | × | × |
| **[66]** | Genetic Algorithm, Fuzzy Controller, Kalman filter. | Energy Saving, User comfort index. | ✓ | ✓ | ✓ | ✓ | ✓ | × | ✓ | ✓ | × | ✓ | × | × | × | × | × | × | ✓ | × | × |
| **[67]** | Artificial Bee Colony, Fuzzy Controllers. | Comfort index, Energy Saving. | × | ✓ | ✓ | ✓ | ✓ | × | ✓ | ✓ | × | ✓ | × | × | × | × | × | × | ✓ | × | × |

| | | | | | | | | | | | | | | | | | | | | |
|---|---|---|---|---|---|---|---|---|---|---|---|---|---|---|---|---|---|---|---|---|
| **[68]** | Genetic Programming, Genetic Algorithm, Fuzzy Logic. | Energy Saving, Comfort index. | ✓ | ✓ | ✓ | ✓ | ✓ | × | ✓ | ✓ | × | ✓ | × | × | × | × | × | × | ✓ | × | × |
| **[69]** | Genetic Algorithm, Kalman Filter, Particle Swarm Optimization (PSO). | Energy Saving, Comfort Index. | ✓ | ✓ | ✓ | ✓ | ✓ | × | ✓ | ✓ | × | ✓ | × | × | × | × | × | × | ✓ | × | ✓ |
| **[70]** | Bat Algorithm, Fuzzy Logic. | Comfort Index, Energy Saving. | × | ✓ | ✓ | ✓ | ✓ | × | ✓ | ✓ | × | ✓ | × | × | × | × | × | × | ✓ | × | × |

Note: Energy Prediction (EP), Energy Optimization (EO), Thermal Comfort (TC), Visual Comfort (VC), Air Quality Comfort (AQC), Sound Comfort (SC), Air Quality (AQ), Temperature (T), Humidity (H), Illumination (Ill), Air Flow (AF), Heat Radiation/Flow (HR/F), Cloth Thermal Insulation (CTI), Metabolic Rate (MR), Water Vapor Pressure (WVP), Single User (SU), Multi-User (MU), Individual Room (IR), External Environment (EE).

## Appendix C

**Table A3.** Energy scheduling techniques and their comparative analysis.

| Ref. | Algorithm / Techniques | Focus Area | Optimization/ Prediction | | Comfort Index Parameters | | | | | | | Pricing Schemes | | | | | Scheduling and Cost Reduction | | | | |
|---|---|---|---|---|---|---|---|---|---|---|---|---|---|---|---|---|---|---|---|---|---|
| | | | FEP | EO | TC | VC | AQC | AQ | T | H | Ill | RTP | DA-RTP | CPP | TOU | FP | APD | OS | R-PAR | RPC | RC |
| **[71]** | Action Dependent Heuristic Dynamic Programming (ADHDP), Neural Network, Backpropagation (BP), Particle Swarm Optimization (PSO). | Reduce Cost, Energy Saving | ✓ | ✓ | × | × | × | × | × | × | × | ✓ | × | × | × | × | × | ✓ | ✓ | ✓ | ✓ |
| **[72]** | Model Predictive Control, Discomfort tolerance index, Hammerstein–Wiener Model, Temperature set-point assignment (TSA) algorithm. | Reduce Cost, Energy Saving | ✓ | ✓ | ✓ | × | × | × | ✓ | × | × | ✓ | × | × | × | × | × | ✓ | × | ✓ | ✓ |
| **[73]** | Genetic Algorithm. | Energy Saving, Reduce Cost, Demand Side Management | × | ✓ | × | × | × | × | × | × | × | × | × | × | × | × | ✓ | ✓ | ✓ | ✓ | ✓ |
| **[74]** | Earthworm Optimization Algorithm, Bacterial Foraging Algorithm. | Energy Saving, Reduce Cost | × | ✓ | × | × | × | × | × | × | × | × | × | ✓ | × | × | ✓ | ✓ | ✓ | ✓ | ✓ |
| **[78]** | Genetic Algorithm, Artificial fish Swarm Optimization (AFSO). | Energy Saving, Reduce Cost, Demand Side | × | ✓ | × | × | × | × | × | × | × | ✓ | × | ✓ | × | × | ✓ | ✓ | ✓ | ✓ | ✓ |
| | | Management | | | | | | | | | | | | | | | | | | | |
| **[79]** | Mixed Integer Linear Programming (MILP), Daily Maximum Energy Scheduling (DMES). | Energy Saving, Reduce Cost | × | ✓ | × | × | × | × | × | × | × | × | × | × | × | × | × | × | × | ✓ | ✓ |
| **[80]** | Sequential quadratic programming, Levenberg–Marquardt, Interior-point. | Energy Saving, Reduce Cost | × | ✓ | ✓ | ✓ | × | × | ✓ | × | × | ✓ | × | × | × | × | ✓ | ✓ | ✓ | ✓ | ✓ |
| **[81]** | Daily Maximum Energy Scheduling (DMES)-Demand Side Management (DSM), Mixed Integer Linear Programming (MILP). | Energy Saving, Reduce Cost | × | ✓ | × | × | × | × | × | × | × | × | × | × | ✓ | × | ✓ | ✓ | ✓ | ✓ | ✓ |
| **[82]** | Dynamic programming algorithm, Regression algorithm, Recurrent Neural Network, Support Vector Regression, Radial Basis Function (RDF) Kernel, Random Forest Regression algorithm. | Energy Saving, Reduce Cost | ✓ | ✓ | ✓ | × | × | × | ✓ | × | × | ✓ | × | × | ✓ | ✓ | × | ✓ | ✓ | ✓ | ✓ |
| **[83]** | Enhanced Differential Evolution (EDE), Teacher Learning-Based Optimization (TLBO). | Reduce Cost, Energy Saving, Demand Side Management | ✓ | ✓ | × | × | × | × | × | × | × | ✓ | ✓ | ✓ | × | × | ✓ | ✓ | ✓ | ✓ | ✓ |
| **[84]** | Model Predictive Control (MPC). | Energy Saving, Reduce Cost | ✓ | ✓ | × | × | × | × | ✓ | × | × | × | × | × | × | × | × | ✓ | × | ✓ | ✓ |
| **[85]** | Genetic Algorithm, Earthworm Optimization Algorithm. | Energy Saving, Reduce Cost, Demand Side Management | × | ✓ | × | × | × | × | × | × | × | × | × | × | ✓ | × | ✓ | ✓ | ✓ | ✓ | ✓ |
| **[86]** | Energy Consumption Scheduling (ECS) Device, Distributed Algorithm. | Reduce Cost, Reduce Peak to Average Ratio | ✓ | ✓ | × | × | × | × | × | × | × | × | × | × | × | × | × | ✓ | ✓ | ✓ | ✓ |
| **[87]** | Dynamic Programming, Genetic Algorithm, Binary Particle Swarm Optimization, Hybrid Scheme GAPSO, Multiple Knapsack Problem (MKP). | Energy Saving, Reduce Cost, Load Management | ✓ | ✓ | × | × | × | × | × | × | × | × | ✓ | ✓ | × | × | ✓ | ✓ | ✓ | ✓ | ✓ |
| **[88]** | Genetic Algorithm (GA), Cuckoo | Energy Saving, | × | ✓ | × | × | × | × | × | × | × | ✓ | × | ✓ | × | × | ✓ | ✓ | ✓ | ✓ | ✓ |

| | | | | | | | | | | | | | | | | | | | | |
|---|---|---|---|---|---|---|---|---|---|---|---|---|---|---|---|---|---|---|---|---|
| | Search Optimization Algorithm (CSOA), Crow Search Algorithm (CSA), smart Electricity Storage System (ESS). | Reduce Cost, Demand Side Management | | | | | | | | | | | | | | | | | | |
| **[89]** | Bacterial Foraging Optimization Algorithm (BFOA), Flower Pollination Algorithm (FPA). | Energy Saving, Reduce Cost, Reduce PAR | × | ✓ | × | × | × | × | × | × | × | ✓ | × | ✓ | × | × | ✓ | ✓ | ✓ | ✓ | ✓ |
| **[90]** | Optimized Home Energy Management System (OHEMS), Genetic Algorithm (GA), Binary Particle Swarm Optimization (BPSO), Wind Driven Optimization (WDO), Bacterial Foraging Optimization (BFO), Hybrid GA-PSO (HGPO) Algorithm, Multiple Knapsack Problem (MKP). | Energy Saving, Reduce Cost, Reduce PAR | ✓ | ✓ | × | × | × | × | ✓ | × | × | × | ✓ | × | × | × | ✓ | ✓ | ✓ | ✓ | ✓ |
| **[91]** | Home Energy Management System (HEMS), Energy Storage System (ESS), Renewable Energy Resources (RES), Earliglow based optimization method, Jaya Algorithm, Enhanced Differential Evolution, Strawberry Algorithm (SBA). | Energy Saving, Reduce Cost, Reduce PAR | × | ✓ | × | × | × | × | × | × | × | × | × | ✓ | ✓ | × | ✓ | ✓ | ✓ | ✓ | ✓ |
| **[92]** | Genetic Harmony Search Algorithm (GHSA), Wind-Driven Optimization (WDO), Harmony Search Algorithm (HSA), Genetic Algorithm (GA). | Energy Saving, Reduce Cost, Reduce PAR | × | ✓ | × | × | × | × | × | × | × | ✓ | × | ✓ | × | × | ✓ | ✓ | ✓ | ✓ | ✓ |
| **[93]** | Genetic Algorithm (GA), Teacher Learning-Based Optimization (TLBO), Linear Programming (LP). | Energy Saving, Reduce Cost, Reduce PAR, user comfort | × | ✓ | × | × | × | × | × | × | × | × | ✓ | × | × | × | ✓ | ✓ | ✓ | ✓ | ✓ |

Note: Future Energy Prediction (FEP), Energy Optimization (EO), Thermal Comfort (TC), Visual Comfort (VC), Air Quality Comfort (AQC), Air Quality (AQ), Temperature (T), Humidity (H), Illumination (Ill), Real Time Pricing (RTP), Day Ahead Real Time Pricing (DA-RTP), Critical Peak Pricing (CPP), Time of Use pricing (TOU), Flat Pricing (FP), Appliance Power Consumption Based Division (APD), Optimal Scheduling (OS), Reduce Peak to Average Ratio (R-PAR), Reduce Power Consumption (RPC), Reduce Cost (RC).